\documentclass[11pt, a4paper]{article}
\usepackage{jheppub, slashed, color,subfig}
\usepackage{enumitem}
\usepackage{multirow}
\usepackage{tikz}
\usepackage{amsmath}
\allowdisplaybreaks[2]
\usepackage{color}
\usepackage[multiple]{footmisc}%for multiple footnotes
\usepackage{textcomp}
\usepackage{amssymb}%for semidirect product symbol
\usetikzlibrary{positioning}
\usetikzlibrary{calc,intersections,through,backgrounds}
\usepackage{amscd}
\usepackage{amsthm}
\usepackage{array} 
\usepackage{graphicx}
\usepackage{adjustbox}
\usepackage{mathrsfs}
\usepackage{dsfont}
\DeclareFontFamily{U}{MnSymbolC}{}
\DeclareSymbolFont{MnSyC}{U}{MnSymbolC}{m}{n}
\DeclareFontShape{U}{MnSymbolC}{m}{n}{
    <-6>  MnSymbolC5
   <6-7>  MnSymbolC6
   <7-8>  MnSymbolC7
   <8-9>  MnSymbolC8
   <9-10> MnSymbolC9
  <10-12> MnSymbolC10
  <12->   MnSymbolC12}{}
\DeclareMathSymbol{\intprod}{\mathbin}{MnSyC}{'270}

\newcommand{\C}{\mathbb{C}}
\newcommand{\CP}{\C P^1}

\newcommand{\Z}{\mathbb{Z}}
\newcommand{\R}{\mathbb{R}}

\newcommand{\til}{\widetilde}

\newcommand{\del}{\partial}

\newcommand{\g}{\mathfrak g}

\newcommand{\zb}{\bar{z}}

\let\nc\newcommand
\let\renc\renewcommand
\nc{\wbar}{\overline}
\let\td\tilde
\let\wtd\widetilde
\let\wht\widehat
\let\mcl\mathcal

\nc{\ab}{{\bar{a}}} \nc{\at}{\tilde{a}} \nc{\ah}{\hat{a}}
\nc{\bb}{{\bar{b}}} 
\nc{\cb}{{\bar{c}}} \nc{\ct}{\tilde{c}} %\nc{\ch}{\hat{c}}
\nc{\fb}{{\bar{f}}} \nc{\ft}{\tilde{f}} \nc{\fh}{\hat{f}}
\nc{\gb}{{\bar{g}}} \nc{\gt}{\tilde{g}} \nc{\gh}{\hat{g}}
\nc{\hb}{{\bar{h}}} \nc{\hh}{\hat{h}} %\nc{\ht}{\tilde{h}}
\nc{\ib}{{\bar{\imath}}} \nc{\ih}{\hat{\imath}} %\nc{\it}{\tilde{\imath}}
\nc{\jb}{{\bar{\jmath}}} \nc{\jt}{\tilde{\jmath}} \nc{\jh}{\hat{\jmath}}
\nc{\kb}{{\bar{k}}} \nc{\kt}{\tilde{k}} \nc{\kh}{\hat{k}}
\nc{\lb}{{\bar{l}}} \nc{\lt}{\tilde{l}} \nc{\lh}{\hat{l}}
\nc{\mb}{{\bar{m}}} \nc{\mt}{\tilde{m}} \nc{\mh}{\hat{m}}
\nc{\nb}{{\bar{n}}} \nc{\nt}{\tilde{n}} \nc{\nh}{\hat{n}}
\nc{\ob}{{\bar{o}}} \nc{\ot}{\tilde{o}} \nc{\oh}{\hat{o}}
\nc{\pb}{{\bar{p}}} \nc{\pt}{\tilde{p}} \nc{\ph}{\hat{p}}
\nc{\qb}{{\bar{q}}} \nc{\qt}{\tilde{q}} \nc{\qh}{\hat{q}}
\nc{\rb}{{\bar{r}}} \nc{\rt}{\tilde{r}} 
\renc{\sb}{{\bar{s}}} \nc{\st}{\tilde{s}} \nc{\sh}{\hat{s}}
\nc{\tb}{{\bar{t}}} \renc{\th}{\hat{t}} %\nc{\tt}{\tilde{t}}
\nc{\ub}{{\bar{u}}} \nc{\ut}{\tilde{u}} \nc{\uh}{\hat{u}}
\nc{\vb}{{\bar{v}}} \nc{\vt}{\tilde{v}} \nc{\vh}{\hat{v}}
\nc{\wb}{{\bar{w}}} \nc{\wt}{\tilde{w}} \nc{\wh}{\hat{w}}
\nc{\xb}{{\bar{x}}} \nc{\xt}{\tilde{x}} \nc{\xh}{\hat{x}}
\nc{\yb}{{\bar{y}}} \nc{\yt}{\tilde{y}} \nc{\yh}{\hat{y}}
\nc{\Ab}{\wbar{A}} \nc{\At}{\wtd{A}} \nc{\Ah}{\wht{A}}
\nc{\Bb}{\wbar{B}} \nc{\Bt}{\wtd{B}} \nc{\Bh}{\wht{B}}
\nc{\Cb}{\wbar{C}} \nc{\Ct}{\wtd{C}} \nc{\Ch}{\wht{C}}
\nc{\Db}{\wbar{D}} \nc{\Dt}{\wtd{D}} \nc{\Dh}{\wht{D}}
\nc{\Eb}{\wbar{E}} \nc{\Et}{\wtd{E}} \nc{\Eh}{\wht{E}}
\nc{\Fb}{\wbar{F}} \nc{\Ft}{\wtd{F}} \nc{\Fh}{\wht{F}}
\nc{\Gb}{\wbar{G}} \nc{\Gt}{\wtd{G}} \nc{\Gh}{\wht{G}}
\nc{\Hb}{\wbar{H}} \nc{\Ht}{\wtd{H}} \nc{\Hh}{\wht{H}}
\nc{\Ib}{\wbar{I}} \nc{\It}{\wtd{I}} \nc{\Ih}{\wht{I}}
\nc{\Jb}{\wbar{J}} \nc{\Jt}{\wtd{J}} \nc{\Jh}{\wht{J}}
\nc{\Kb}{\wbar{K}} \nc{\Kt}{\wtd{K}} \nc{\Kh}{\wht{K}}
\nc{\Lb}{\wbar{L}} \nc{\Lt}{\wtd{L}} \nc{\Lh}{\wht{L}}
\nc{\Mb}{\wbar{M}} \nc{\Mt}{\wtd{M}} \nc{\Mh}{\wht{M}}
\nc{\Nb}{\wbar{N}} \nc{\Nt}{\wtd{N}} \nc{\Nh}{\wht{N}}
\nc{\Ob}{\wbar{O}} \nc{\Ot}{\wtd{O}} \nc{\Oh}{\wht{O}}
\nc{\Pb}{\wbar{P}} \nc{\Pt}{\wtd{P}} \nc{\Ph}{\wht{P}}
\nc{\Qb}{\wbar{Q}} \nc{\Qt}{\wtd{Q}} \nc{\Qh}{\wht{Q}}
\nc{\Rb}{\wbar{R}} \nc{\Rt}{\wtd{R}} \nc{\Rh}{\wht{R}}
\nc{\Sb}{\wbar{S}} \nc{\St}{\wtd{S}} \nc{\Sh}{\wht{S}}
\nc{\Tb}{\wbar{T}} \nc{\Tt}{\wtd{T}} \nc{\Th}{\wht{T}}
\nc{\Ub}{\wbar{U}} \nc{\Ut}{\wtd{U}} \nc{\Uh}{\wht{U}}
\nc{\Vb}{\wbar{V}} \nc{\Vt}{\wtd{V}} \nc{\Vh}{\wht{V}}
\nc{\Wb}{\wbar{W}} \nc{\Wt}{\wtd{W}} \nc{\Wh}{\wht{W}}
\nc{\Xb}{\wbar{X}} \nc{\Xt}{\wtd{X}} \nc{\Xh}{\wht{X}}
\nc{\Yb}{\wbar{Y}} \nc{\Yt}{\wtd{Y}} \nc{\Yh}{\wht{Y}}
\nc{\Zb}{\wbar{Z}} \nc{\Zt}{\wtd{Z}} \nc{\Zh}{\wht{Z}}

\nc{\CA}{\mcl{A}} \nc{\CAb}{\wbar{\CA}} \nc{\CAt}{\wtd{\CA}} \nc{\CAh}{\wht{\CA}}
\nc{\CB}{\mcl{B}} \nc{\CBb}{\wbar{\CB}} \nc{\CBt}{\wtd{\CB}} \nc{\CBh}{\wht{\CB}}
\nc{\cD}{\mcl{D}} \nc{\cDb}{\wbar{\cD}} \nc{\cDt}{\wtd{\cC}} \nc{\cDh}{\wht{\cD}}
\nc{\CE}{\mcl{E}} \nc{\CEb}{\wbar{\CE}} \nc{\CEt}{\wtd{\CE}} \nc{\CEh}{\wht{\CE}}
\nc{\CF}{\mcl{F}} \nc{\CFb}{\wbar{\CF}} \nc{\CFt}{\wtd{\CF}} \nc{\CFh}{\wht{\CF}}
\nc{\CG}{\mcl{G}} \nc{\CGb}{\wbar{\CG}} \nc{\CGt}{\wtd{\CG}} \nc{\CGh}{\wht{\CG}}
\nc{\CH}{\mcl{H}} \nc{\CHb}{\wbar{\CH}} \nc{\CHt}{\wtd{\CH}} \nc{\CHh}{\wht{\CH}}
\nc{\CI}{\mcl{I}} \nc{\CIb}{\wbar{\CI}} \nc{\CIt}{\wtd{\CI}} \nc{\CIh}{\wht{\CI}}
\nc{\CJ}{\mcl{J}} \nc{\CJb}{\wbar{\CJ}} \nc{\CJt}{\wtd{\CJ}} \nc{\CJh}{\wht{\CJ}}
\nc{\CK}{\mcl{K}} \nc{\CKb}{\wbar{\CK}} \nc{\CKt}{\wtd{\CK}} \nc{\CKh}{\wht{\CK}}
\nc{\CL}{\mcl{L}} \nc{\CLb}{\wbar{\CL}} \nc{\CLt}{\wtd{\CL}} \nc{\CLh}{\wht{\CL}}
\nc{\CM}{\mcl{M}} \nc{\CMb}{\wbar{\CM}} \nc{\CMt}{\wtd{\CM}} \nc{\CMh}{\wht{\CM}}
\nc{\CN}{\mcl{N}} \nc{\CNb}{\wbar{\CN}} \nc{\CNt}{\wtd{\CN}} \nc{\CNh}{\wht{\CN}}
\nc{\CO}{\mcl{O}} \nc{\COb}{\wbar{\CO}} \nc{\COt}{\wtd{\CO}} \nc{\COh}{\wht{\CO}}
\nc{\CQ}{\mcl{Q}} \nc{\CQb}{\wbar{\CQ}} \nc{\CQt}{\wtd{\CQ}} \nc{\CQh}{\wht{\CQ}}
\nc{\CR}{\mcl{R}} \nc{\CRb}{\wbar{\CR}} \nc{\CRt}{\wtd{\CR}} \nc{\CRh}{\wht{\CR}}
\nc{\CS}{\mcl{S}} \nc{\CSb}{\wbar{\CS}} \nc{\CSt}{\wtd{\CS}} \nc{\CSh}{\wht{\CS}}
\nc{\CT}{\mcl{T}} \nc{\CTb}{\wbar{\CT}} \nc{\CTt}{\wtd{\CT}} \nc{\CTh}{\wht{\CT}}
\nc{\CU}{\mcl{U}} \nc{\CUb}{\wbar{\CU}} \nc{\CUt}{\wtd{\CU}} \nc{\CUh}{\wht{\CU}}
\nc{\CV}{\mcl{V}} \nc{\CVb}{\wbar{\CV}} \nc{\CVt}{\wtd{\CV}} \nc{\CVh}{\wht{\CV}}
\nc{\CW}{\mcl{W}} \nc{\CWb}{\wbar{\CW}} \nc{\CWt}{\wtd{\CW}} \nc{\CWh}{\wht{\CW}}
\nc{\CX}{\mcl{X}} \nc{\CXb}{\wbar{\CX}} \nc{\CXt}{\wtd{\CX}} \nc{\CXh}{\wht{\CX}}
\nc{\CY}{\mcl{Y}} \nc{\CYb}{\wbar{\CY}} \nc{\CYt}{\wtd{\CY}} \nc{\CYh}{\wht{\CY}}
\nc{\CZ}{\mcl{Z}} \nc{\CZb}{\wbar{\CZ}} \nc{\CZt}{\wtd{\CZ}} \nc{\CZh}{\wht{\CZ}}

\let\eps\epsilon
\let\ups\upsilon

\let\veps\varepsilon
\let\vtht\vartheta
\let\vsgm\varsigma
\let\vphi\varphi
\let\vrho\varrho

\nc{\alphab}{\bar{\alpha}} \nc{\alphat}{\td{\alpha}} \nc{\alphah}{\hat{\alpha}}
\nc{\betab}{\bar{\beta}}   \nc{\betat}{\td{\beta}}   \nc{\betah}{\hat{\beta}} 
\nc{\gammab}{\bar{\gamma}} \nc{\gammat}{\td{\gamma}} \nc{\gammah}{\hat{\gamma}} 
\nc{\deltab}{\bar{\delta}} \nc{\deltat}{\td{\delta}} \nc{\deltah}{\hat{\delta}} 
\nc{\epsilonb}{\bar{\eps}} \nc{\epsilont}{\td{\eps}} \nc{\epsilonh}{\hat{\eps}} 
\nc{\vepsb}{\bar{\veps}}   \nc{\vepst}{\td{\veps}}   \nc{\vepsh}{\hat{\veps}} 
\nc{\zetab}{\bar{\zeta}}   \nc{\zetat}{\td{\zeta}}   \nc{\zetah}{\hat{\zeta}} 
\nc{\etab}{\bar{\eta}}     \nc{\etat}{\td{\eta}}     \nc{\etah}{\hat{\eta}} 
\nc{\thetab}{\bar{\theta}} \nc{\thetat}{\td{\theta}} \nc{\thetah}{\hat{\theta}} 
\nc{\vthetab}{\bar{\vtht}} \nc{\vthetat}{\td{\vtht}} \nc{\vthetah}{\hat{\vtht}} 
\nc{\lambdab}{\bar{\lambda}} \nc{\lambdat}{\td{\lambda}} \nc{\lambdah}{\hat{\lambda}} 
\nc{\iotab}{\bar{\iota}}   \nc{\iotat}{\td{\iota}}   \nc{\iotah}{\hat{\iota}} 
\nc{\kappab}{\bar{\kappa}} \nc{\kappat}{\td{\kappa}} \nc{\kappah}{\hat{\kappa}} 
\nc{\lmdb}{\bar{\lmd}}     \nc{\lmdt}{\td{\lmd}}     \nc{\lmdh}{\hat{\lmd}} 
\nc{\mub}{\bar{\mu}}       \nc{\mut}{\td{\mu}}       \nc{\muh}{\hat{\mu}} 
\nc{\nub}{\bar{\nu}}       \nc{\nut}{\td{\nu}}       \nc{\nuh}{\hat{\nu}} 
\nc{\xib}{\bar{\xi}}       \nc{\xit}{\td{\xi}}       \nc{\xih}{\hat{\xi}} 
\nc{\pib}{\bar{\pi}}       \nc{\pit}{\td{\pi}}       \nc{\pih}{\hat{\pi}} 
\nc{\vpib}{\bar{\vpi}}     \nc{\vpit}{\td{\vpi}}     \nc{\vpih}{\hat{\vpi}} 
\nc{\rhob}{\bar{\rho}}     \nc{\rhot}{\td{\rho}}     \nc{\rhoh}{\hat{\rho}} 
\nc{\vrhob}{\bar{\vrho}}   \nc{\vrhot}{\td{\vrho}}   \nc{\vrhoh}{\hat{\vrho}} 
\nc{\sigmab}{\bar{\sigma}} \nc{\sigmat}{\td{\sigma}} \nc{\sigmah}{\hat{\sigma}} 
\nc{\vsigmab}{\bar{\vsgm}} \nc{\vsigmat}{\td{\vsgm}} \nc{\vsigmah}{\hat{\vsgm}} 
\nc{\taub}{\bar{\tau}}     \nc{\taut}{\td{\tau}}     \nc{\tauh}{\hat{\tau}} 
\nc{\upsb}{\bar{\ups}} \nc{\upst}{\td{\ups}} \nc{\upsh}{\hat{\ups}} 
\nc{\phib}{\bar{\phi}}     \nc{\phit}{\td{\phi}}     \nc{\phih}{\hat{\phi}} 
\nc{\varphib}{\bar{\vphi}}   \nc{\varphit}{\td{\vphi}}   \nc{\varphih}{\hat{\vphi}} 
\nc{\chib}{\bar{\chi}}     \nc{\chit}{\td{\chi}}     \nc{\chih}{\hat{\chi}} 
\nc{\omegab}{\bar{\omega}} \nc{\omegat}{\td{\omega}} \nc{\omegah}{\hat{\omega}} 

\nc{\Gammab}{\wbar{\Gamma}}     \nc{\Gammat}{\wtd{\Gamma}}     \nc{\Gammah}{\wht{\Gamma}}
\nc{\Deltab}{\wbar{\Delta}}     \nc{\Deltat}{\wtd{\Delta}}     \nc{\Deltah}{\wht{\Delta}}
\nc{\Thetab}{\wbar{\Theta}}     \nc{\Thetat}{\wtd{\Theta}}     \nc{\Thetah}{\wht{\Theta}}
\nc{\Lambdab}{\wbar{\Lambda}}   \nc{\Lambdat}{\wtd{\Lambda}}   \nc{\Lambdah}{\wht{\Lambda}}
\nc{\Xib}{\wbar{\Xi}}           \nc{\Xit}{\wtd{\Xi}}           \nc{\Xih}{\wht{\Xi}}
\nc{\Pib}{\wbar{\Pi}}           \nc{\Pit}{\wtd{\Pi}}           \nc{\Pih}{\wht{\Pi}}
\nc{\Sigmab}{\wbar{\Sigma}}     \nc{\Sigmat}{\wtd{\Sigma}}     \nc{\Sigmah}{\wht{\Sigma}}
\nc{\Upsilonb}{\wbar{\Upsilon}} \nc{\Upsilont}{\wtd{\Upsilon}} \nc{\Upsilonh}{\wht{\Upsilon}}
\nc{\Phib}{\wbar{\Phi}}         \nc{\Phit}{\wtd{\Phi}}         \nc{\Phih}{\wht{\Phi}}
\nc{\Psib}{\wbar{\Psi}}         \nc{\Psit}{\wtd{\Psi}}         \nc{\Psih}{\wht{\Psi}}
\nc{\Omegab}{\wbar{\Omega}}     \nc{\Omegat}{\wtd{\Omega}}     \nc{\Omegah}{\wht{\Omega}}

\nc{\txd}{d}

\newcommand{\cA}{{\cal A}}

\newcommand{\cF}{{\cal F}}

\newcommand{\Tr}{~\textrm{Tr}}
\newcommand{\inv}{^{-1}}
\newcommand{\dl}{\delta}
\nc{\tcos}{\textrm{cos}}
\nc{\tsin}{\textrm{sin}}

\nc{\al}{\alpha}
\nc{\bt}{\beta}
\nc{\gm}{\gamma}
\nc{\rh}{\rho}
\nc{\zt}{\zeta}
\nc{\Dl}{\Delta}
\nc{\TD}{\til{\Delta}}

\nc{\sg}{\Sigma}

\nc{\rd}{0.75}
\begin{document}

%Moreover, we argue that that T-duality with the D5-NS5 brane system gives rise to categorification of the %relevant
%Yang-Baxter equation. %corresponding lattice models.
% We also describe how Costello's theory with supergroup gauge symmetry can arise from certain D4-NS5 brane configurations.
%OR The obstruction to the mathematical
%\newpage
	%\tableofcontents

\title{Integrable lattice models and holography}
\author{Meer Ashwinkumar}
%\author[a,b,1]{Second Author%
%\note{On leave from XXX.}}
%\author[a,c]{and Third Author}
\affiliation{Department of Physics, National University of Singapore,\\
2 Science Drive 3, 117551 Singapore}
%\affiliation[b]{Department, University,\\
%Street number, City, Country}
%\affiliation[c]{Another University,\\
%Street number, City, Country}
\emailAdd{meerashwinkumar@u.nus.edu}
%\emailAdd{buthor@univ.country}
%\emailAdd{cuthor@another.univ.country}
\abstract{We study four-dimensional Chern-Simons theory %on a manifold with boundary
%product of a disk and a Riemann surface
on $D\times \C$ (where $D$ is a disk), which is understood to describe rational solutions of the Yang-Baxter equation from the work of Costello, Witten and Yamazaki. We find that the theory is dual to a boundary theory, that is a three-dimensional analogue of the two-dimensional chiral WZW model. %We derive
This boundary theory gives rise to a current algebra that turns out to be an “analytically-continued” toroidal Lie algebra. In addition, we show how certain bulk correlation functions of two and three Wilson lines can be captured by boundary correlation functions of local operators in the three-dimensional WZW model. In particular, we reproduce %Costello, Witten and Yamazaki’s result for 
the leading and subleading nontrivial contributions to the rational $R$-matrix  %order in $\hbar$ 
purely from the boundary theory. }
%\keywords{Chern-Simons Theories, Lattice Integrable Models}
%\arxivnumber{2003.08931}
%\begin{document}
\dedicated{Dedicated to my mother, Dr. Indra.}
\maketitle

%\maketitle
\noindent
\section{Introduction}
A relatively novel approach to the study of %the 
integrable lattice models % of classical statistical mechanics, 
underlaid by the Yang-Baxter equation with spectral parameters is that of four-dimensional Chern-Simons theory, first proposed by Costello \cite{Costello1,Costello2}, and subsequently studied by Costello, Witten and Yamazaki \cite{CWY,CWY2,CWY3} in depth. 

The theory is defined by the path integral involving the classical action 
\begin{equation}
S=\frac{1}{2\pi\hbar}\int_{ \Sigma\times C} \omega \wedge \textrm{Tr}\bigg(\cA\wedge d\cA + \frac{2}{3} \cA\wedge \cA\wedge \cA\bigg),
\end{equation} 
where $\cA$ is a complex-valued gauge field, $\Sigma$ is a 2-manifold,
and $C$ is a Riemann surface endowed with a holomorphic
one-form $\omega = \omega(z)dz$. It is topological along $\sg$ (modulo a framing anomaly), but has holomorphic dependence on $C$, and %can be shown to be invariant under 
moreover has a complex gauge group, $G$. 
As shown in \cite{CWY,CWY2}, the nontrivial operators of the theory are Wilson lines, whose correlation functions  realize the Yang-Baxter equation with spectral parameters, as well as the underlying Yangian algebra, quantum affine algebra, and elliptic quantum group of its rational, trigonometric and elliptic solutions, respectively. Furthermore, the boundary Yang-Baxter equation can likewise be realized by studying 4d Chern-Simons theory on an orbifold \cite{BS,BS2}. 
%It has a complex gauge group, denoted G.

%Explain theory and what it describes/algebras it realizes. 

For $C=\R \times S^1$, the 4d Chern-Simons action can be dimensionally reduced along $S^1$ to that of 3d analytically-continued Chern-Simons theory. In fact, the corresponding quantum field theories have been shown to be T-dual \cite{yamazaki}.
%T-dual to 3d analytically continued CS. 
However, unlike ordinary 3d Chern-Simons theory, much of the work on 4d Chern-Simons relied on the path integral and Feynman diagrams alone, and no use was made of canonical quantization or holography. %, unlike 3d Chern-Simons theory.
%which is equivalently determined via the 2d chiral WZW model. 
This was due to the infrared-free nature of 4d Chern-Simons, whereby it was straightforward to deduce a local procedure to compute the expectation values of Wilson line configurations of interest. 

%Must write action as well as rational $R$-matrix at order hbar and argue for its completion.

%Point out that the path integral of the theory is only perturbatively defined%

Nevertheless, given the importance of the 2d chiral Wess-Zumino-Witten (WZW) model dual to 3d Chern-Simons theory as a straightforward example of a holographic dual, %as well as 
and for describing edge modes of the nonabelian fractional quantum Hall effect, 
%(see Tong's notes, which shows generalized chiral WZW action and also mentions braiding of wilson lines/conformal blocks. also see Moore-Read and Cabra,Fradkin et al. which discusses cosets),
 it is of interest to investigate the existence of a holographic dual of 4d Chern-Simons theory.
%3d analogue of the WZW model.
In this work, we shall indeed derive such a dual boundary theory for 4d Chern-Simons on $D\times C$ (where $D$ is the disk), which turns out to be a three-dimensional analogue of the 2d chiral WZW model. 

We shall focus on the boundary dual of 4d Chern-Simons with $C=\C$, % which in realizes
which is known to give rise to rational solutions of the Yang-Baxter equation \cite{CWY}. These $R$-matrices are intertwining operators for representations of the Yangian algebra, and thus the classical integrable lattice models of concern are equivalent to Heisenberg XXX quantum spin chains. For example, one such lattice model is the rational six-vertex model, which is equivalent to the XXX$_{1/2}$ spin chain. As we shall see, the 3d ``chiral" WZW model we derive furnishes an alternative and convenient method for computing the rational $R$-matrices explicitly. 

%Rational six-vertex model - equivalent to XXX1/2 spin chain, XXX spin chains in general underlaid by yangian algebra. 

Given a 3d analogue of the 2d chiral WZW model, the first natural question to ask is if it admits a current algebra analogous to an affine Kac-Moody algebra. In Section 2, we shall show that this is indeed the case, i.e., the 3d ``chiral" WZW model furnishes a particular limit of an ``analytically-continued" toroidal Lie algebra. 

One 
%also may wonder if
would also like to verify that the boundary theory captures the correlation functions of the bulk theory. %For instance, the $R$-matrix has been computed in the bulk theory to be 
For instance, the bulk correlator of two straight, perpendicular Wilson lines along $\sg$, at points $z_1$ and $z_2$ on $\C$, and in representations $R_1$ and $R_2$ of the generators of $G$, realizes the $R$-matrix, i.e., %To be precise, 
it is computed to be
\begin{equation}
\til{R}_{12}(z_1-z_2)=\mathds{1}+\frac{\hbar}{z_1-z_2}T_{R_1}^a\otimes T_{ R_2 a}+{O}(\hbar^2)
\end{equation}
to linear order in $\hbar$, with the full expression for the $R$-matrix following from general theorems %(up to a change of variables) 
\cite{Drinfeld,Chari}. It would be satisfying to obtain this result solely from the boundary theory, and indeed, this is what we do in Section 3 by evaluating a four-point function of local boundary operators. 
Furthermore, via the boundary theory, we demonstrate the topological invariance along $D$ of the bulk correlator at order $\hbar$. %Finally, w

We also explain how one may explicitly derive higher order contributions to the $R$-matrix, and demonstrate this by computing the $\hbar^2$ contribution to the correlation function of %perpendicular  
two straight, perpendicular Wilson lines explicitly. Subsequently, we show that the boundary theory also reproduces the order $\hbar$ correlation function of a pair of parallel Wilson lines. Finally, we consider three Wilson lines, all crossing each other, and show, modulo the framing anomaly, that their order $\hbar^2$ correlation function is reproduced by a six-point function in the boundary theory.

%This work has two main parts, firstly we derive a current algebra which turns out to be... Secondly we derive the hbar contribution above. Write bulk to boundary operator equivalence and cite fradkin while mentioning its relevance to nonabelian fractional quantum Hall effect described by ordinary 3d Chern-Simons theory. Moreover we show that the topological invariance of the bulk theory is preserved in the boundary theory. Finally, we also derive the hbar2 contribution to the $R$-matrix.
\section{3d ``Chiral" WZW Model}
Let us start with 4d Chern-Simons theory with complex gauge group, $G$, defined on $\Sigma\times C$, where $\Sigma$ is a disk, denoted $D$, and $C$ is the complex plane, $\C$. Its action is %or $\C/(\Z+\tau \Z)$, i.e.,
\begin{equation}\label{introaction}
S=\frac{1}{2\pi\hbar}\int_{ D\times \C} dz\wedge \textrm{Tr}\bigg(\cA\wedge d\cA + \frac{2}{3} \cA\wedge \cA\wedge \cA\bigg).
\end{equation} 
Here, $\cA$ can be understood to be the partial connection
\begin{equation}
\cA=\cA_rdr + \cA_\vphi d\vphi +\cA_{\zb} d\zb,
\end{equation}
where $(r,\vphi)$ are polar coordinates on $D$ and $(z,\zb)$ are complex coordinates on $\C$.

Let us first vary the action to find the equations of motion. Doing so, one finds
\begin{equation}
\delta S=\frac{1}{2\pi\hbar}\int_{ D\times \C} dz\wedge \textrm{Tr}\bigg(\delta \cA \wedge \cF+ d(\delta \cA\wedge \cA)\bigg).
\end{equation}
The second term of the variation is a boundary term via Stoke's theorem. In order 
to ensure that we have equations of motion free from boundary corrections, we shall impose the boundary condition $\cA_{\zb}=0$, whereupon the boundary term vanishes.

This boundary condition is also necessary to achieve gauge invariance in the presence of boundaries. It can be shown that \eqref{introaction} is equivalent to 
\begin{equation}\label{gaugeinva}
S=-\frac{1}{2\pi\hbar}\int_{ D\times \C} z \textrm{Tr}\bigg(F\wedge F\bigg)+\frac{1}{2\pi\hbar}\int_{ \del D\times \C} z\textrm{Tr}\bigg(\cA\wedge d\cA + \frac{2}{3} \cA\wedge \cA\wedge \cA\bigg),
\end{equation}
where $\cA$ has been extended to a \textit{full} connection over $D\times \C$, i.e., $\cA=\cA_rdr + \cA_\vphi d\vphi +\cA_{z} dz +\cA_{\zb} d\zb$. The boundary term on the RHS of \eqref{gaugeinva} depends only on the components $\cA_{\vphi}$, $\cA_{z}$ and $\cA_{\zb}$, and vanishes using the boundary conditions $\cA_{\zb}=0$ and $\cA_{z}=0$. The remaining term is gauge invariant under large gauge transformations, i.e.,
\begin{equation}\label{gaugetr}
\cA\rightarrow U\cA U^{-1}-dUU^{-1}.
\end{equation}
However, we ought to restrict $U$ such that the boundary conditions $\cA_{\zb}=\cA_{z}=0$ are preserved. We shall achieve this by insisting that $U$ tends to the identity element of $G$ at the boundary.

Now, having imposed the boundary condition $\cA_{\zb}=0$, the action \eqref{introaction} is equivalent to
\begin{equation}\label{2ndaction}
\frac{1}{2\pi\hbar}\int dz\wedge dr \wedge d\vphi \wedge d\zb ~\textrm{Tr} \bigg(2\cA_{\zb}\cF_{r\vphi}-\cA_r\del_{\zb}\cA_{\vphi}+\cA_{\vphi}\del_{\zb}\cA_r\bigg),
\end{equation} 
upon integration by parts. Varying the Lagrange multiplier field, $\cA_{\zb}$, implements the constraint
$\cF_{r\vphi}=0$, which is solved by 
\begin{equation}\label{sols}
\begin{aligned}
\cA_r=-\del_rg g^{-1}, ~~
\cA_\vphi=-\del_\vphi g g^{-1},
\end{aligned}
\end{equation}
where $g$ is a %single-valued (single-valued requires simply connected G)
map $g:D\times \C\rightarrow G$.

Changing variables from $\cA_r$ and $\cA_\vphi$ to $g$ in the functional integral, we note that, just as in 3d Chern-Simons theory \cite{Elitzur}, no Jacobian appears when transforming the measure, i.e.,
\begin{equation}
\frac{1}{vol~G }\int D\cA_r D\cA_{\vphi} ~\delta(\cF_{r\vphi}) = \frac{1}{vol~G }\int Dg,
\end{equation}
where the expression on the RHS is the relevant Haar measure, divided by the volume of the gauge group. %defined below \eqref{gaugetr}. 
Furthermore, substituting the solutions \eqref{sols} into \eqref{2ndaction}, we obtain the action
\begin{equation}\label{cwzw3}
S(g)=\frac{1}{2\pi\hbar}\int_{S^1\times \C}d\vphi\wedge dz\wedge d\zb ~\textrm{Tr} (\del_{\vphi}gg^{-1}\del_{\zb} gg\inv)+\frac{1}{6\pi\hbar}\int_{D\times\C} dz\wedge\Tr(dg g\inv \wedge dg g\inv \wedge  dg g\inv),
\end{equation}
which takes the form of a three-dimensional analogue of the 2d chiral WZW model. Now, a large gauge transformation \eqref{gaugetr} amounts to $g\rightarrow Ug$ in \eqref{sols}. As a result, we may change the value of $g$ in the interior without changing its value at the boundary, so \eqref{cwzw3} only depends on $g$ at the boundary. Note that in the 2d WZW model %arising from
dual to 3d Chern-Simons theory on $D\times \R$ with a compact gauge group studied in \cite{Elitzur}, the fact that the Wess-Zumino term does not depend on the choice of extension of $g$ over a 3-manifold requires quantization of the WZW level/Chern-Simons coupling. However, the coupling $\hbar$ that appears in \eqref{cwzw3} is not quantized. This difference stems from the fact that invariance of the 3d Chern-Simons action under large gauge transformations of the form \eqref{gaugetr} (which leads to the Wess-Zumino term being independent of the choice of extension) requires such a quantization, while large gauge invariance of the 4d Chern-Simons action does not require such a quantization, as shown below \eqref{gaugeinva}.
The fact that \eqref{cwzw3} only depends on $g$ at the boundary implies that we can divide out the volume of the gauge group to obtain the path integral 
\begin{equation}
\int Dg~ e ^{iS(g)},
\end{equation}
where $g$ is now the %single-valued 
map $g:\del D \times \C \rightarrow G$.

Varying the action \eqref{cwzw3} gives us
\begin{equation}\label{vari}
\delta S= -\frac{1}{\pi\hbar}\int d\vphi\wedge dz \wedge d\zb \Tr (g\inv\delta g \del_{\vphi}(g\inv\del_{\zb} g)),
\end{equation}
whereby we obtain the classical equation of motion 
\begin{equation}
\del_{\vphi}(g\inv\del_{\zb} g)=0,
\end{equation}
which is equivalent to $\del_{\zb}(\del_{\vphi}g g\inv)=0$, and is solved by
\begin{equation}\label{eomsol}
g(z,\zb,\varphi)=A(z,\varphi)B(z,\zb).
\end{equation}
 The equations $\del_{\vphi}(g\inv\del_{\zb} g)=0$ and $\del_{\zb}(\del_{\vphi}g g\inv)=0$ are in fact equivalent to the current conservation equations for the symmetry of the action under the transformation 
\begin{equation}\label{gxg}
g(\vphi,z,\zb)\rightarrow \tilde{\Omega}(\vphi,z)g \Omega(z,\zb),
\end{equation}
where $\tilde{\Omega}$ and $\Omega$ give rise to the conserved currents $J_\vphi=-\frac{1}{\pi\hbar}\del_{\vphi}gg\inv$ and $J_{\zb}=-\frac{1}{\pi\hbar}g\inv\del_{\zb}g$, respectively.

\subsection{Current Algebra via Canonical Quantization}
We are now interested in computing a Poisson bracket involving the expression $J_\vphi=-\frac{1}{\pi\hbar}\del_{\vphi}gg\inv$, which we shall eventually use to obtain a quantum mechanical commutation relation in the form of a current algebra. %In what follows, we shall analytically continue $z$ and $\zb$ such that they take real values, and  
In what follows, we shall take $\zb$ to be the (complex) time direction.

Now, given an arbitrary action that is first order in the time derivative with dynamical variables $\phi^i$, i.e., 
\begin{equation}\label{genact}
I=\int dt \mathscr{A}(\phi)\frac{d\phi^i}{dt}, 
\end{equation}
its variation takes the form 
\begin{equation}
\delta I=\int dt~ \omega_{ij}\delta\phi^i \frac{d\phi^j}{dt},
\end{equation}
where $\omega_{ij}=\frac{\del}{\del\phi^i} \mathscr{A}_j-\frac{\del}{\del\phi^j} \mathscr{A}_i$ is the symplectic structure on the classical phase space. The Poisson bracket of any two functions $X$ and $Y$ on the phase space is then defined by 
\begin{equation}
[X,Y]_{PB}=\omega^{ij}\frac{\del X}{\del \phi^i}\frac{\del Y}{\del \phi^j},
\end{equation}
where $\omega^{jk}\omega_{kl}=\delta^j_l$ \cite{w}.

 For the 3d ``chiral" WZW model, %we are interested in, 
 the variation \eqref{vari} implies that its phase space symplectic structure is given by $\omega = 1_{\mathfrak{g}} \otimes \frac{(-1)}{\pi\hbar}\frac{\del}{\del\vphi}\otimes 1_z$, where $1_{\mathfrak{g}}$ acts on the Lie algebra index, $\frac{(-1)}{\pi\hbar}\frac{\del}{\del\vphi}$ acts on the $\vphi$ coordinate, and $1_z$ acts on the $z$ coordinate. Its inverse is therefore
 \begin{equation}
\omega\inv= 1_{\mathfrak{g}}\otimes (-\pi\hbar)\bigg(\frac{\del}{\del\vphi}\bigg)\inv\otimes 1_z .
 \end{equation}
 
 Let us now compute the Poisson bracket of $X=\Tr A \frac{\del g}{\del \vphi}g\inv(\vphi,z)$ and $Y=\Tr B \frac{\del g}{\del \vphi'}g\inv(\vphi',z')$, where $A$ and $B$ are arbitrary generators of the group $G$. In the notation of \eqref{genact}, this can be done by evaluating $\delta X \delta Y =\frac{\del X}{\del \phi^i}\frac{\del Y}{\del \phi^j}\delta \phi^i\delta\phi^j$, and subsequently replacing $\dl\phi^i\dl\phi^j$ by $\omega^{ij}$. Proceeding in this manner, we find
 \begin{equation}\label{xyvar}
 \delta X \delta Y=\Tr ~g\inv (\vphi,z)A g(\vphi,z)\frac{\del}{\del\vphi}(g\inv\dl g (\vphi,z))\cdot\Tr ~g\inv (\vphi',z')B g(\vphi',z')\frac{\del}{\del\vphi'}(g\inv\dl g (\vphi',z')). 
 \end{equation}
 To obtain the Poisson bracket, we ought to replace $(g\inv \dl g(\vphi,z))^a(g\inv \dl g(\vphi',z'))^b$ (where $a$ and $b$ are Lie algebra indices) by 
 \begin{equation}
 \delta^{ab}(-\pi\hbar)\theta(\vphi-\vphi')\dl(z-z'),
 \end{equation}
 where $\theta(\vphi-\vphi')$ is an inverse of $\frac{\del}{\del \vphi} $.
 %\footnote{The formal delta function in $z$-space can be understood as the usual delta function with contour integral passing through the support of the delta function, or via Cauchy's integral formula where the contour contains the support of the delta function. See stackexchange} 
 Therefore, $\frac{\del}{\del \vphi}(g\inv \dl g(\vphi,z))^a\cdot\frac{\del}{\del \vphi'}(g\inv \dl g(\vphi',z'))^b$ in \eqref{xyvar} ought to be replaced by $\dl^{ab}\pi\hbar\dl'(\vphi-\vphi')\dl(z-z')$.
 Hence, we arrive at the Poisson bracket
 \begin{equation}\label{pbrckt}
 \begin{aligned}
 \left[X,Y\right]_{PB}&=\pi\hbar\dl'(\vphi-\vphi')\dl(z-z')\Tr~g\inv(\vphi,z)Ag(\vphi,z)g\inv(\vphi',z')Bg(\vphi',z')
 \\&=\pi\hbar\dl(\vphi-\vphi')\dl(z-z')\Tr\bigg([A,B]\frac{\del g}{\del\vphi}g\inv\bigg)+\pi\hbar\dl'(\vphi-\vphi')\dl(z-z')\Tr~AB.
 \end{aligned}
 \end{equation}
 Upon rescaling both sides of this equation by $\big(\frac{1}{-\pi \hbar}\big)^2$, we arrive at a classical current algebra. %Also note that the term that contains $\hbar$, which takes the form of a central extension, arises from a central extension term that appears in the classical Poisson bracket \eqref{pbrckt}. 
 Note that we have found a central extension term at the classical level. This is analogous to the case of the standard 2d WZW model 
with kinetic term in ``chiral'' form, i.e., using lightcone coordinates, wherein such a central extension term also appears in the classical Poisson bracket involving currents when taking one of the lightcone directions to be time, as shown in \cite{w}.

 In the quantum theory, the Poisson bracket \eqref{pbrckt} corresponds to a canonical commutation relation that takes the form
\begin{equation} \label{firstquantum}
  \left[X,Y\right]=-i\pi\hbar\dl(\vphi-\vphi')\dl(z-z')\Tr\bigg([A,B]\frac{\del g}{\del\vphi}g\inv\bigg)-i\pi\hbar\dl'(\vphi-\vphi')\dl(z-z')\Tr~AB + \ldots.
\end{equation}
Note that $\hbar$ here is a coupling, based on the definition of the action in \eqref{introaction}, and we have replaced the Poisson bracket $\left[X,Y\right]_{PB}$ by $i\left[X,Y\right]$. 
%Here, Planck's constant has
If we instead set $\hbar$ to 1 in the action and replace the Poisson bracket $\left[X,Y\right]_{PB}$ by $\frac{i}{\til{\hbar}}\left[X,Y\right]$, where $\til{\hbar}$ is Planck's constant, then we will obtain \eqref{firstquantum} with $\hbar$ replaced by $\til{\hbar}$. The terms represented by the ellipsis in \eqref{firstquantum} indicate possible higher order terms in Planck's constant. 

As we shall explain now, 
such quantum corrections to the classical current algebra are expected to be fixed by a symmetry of the quantum 3d WZW model.
%expected to take the form of total derivatives.
% that vanish when integrated over $\vphi$ and $z$. 
Firstly, for $C=\C$, the bulk 4d Chern-Simons theory computes the rational R-matrix, that has $G$ as an automorphism group, and therefore its dual quantum 3d WZW model ought to have $G$ as a symmetry. 
%In \cite{CWY}, the quantization of the bulk theory involved dividing only by gauge transformations that are $\mathds{1}$ at infinity along $\C$,  leaving constant gauge transformations at infinity along $\C$ that behave as global symmetries, which corresponded to the automorphism group of the R-matrix. In our case, we also have a boundary for $\Sigma=D$, and the gauge transformations that we divide by tend to $\mathds{1}$ at this boundary, implying that the gauge transformations we are left with depend on $\vphi$ and $z$ at this boundary (the boundary condition $\cA_z=0$ is not violated since there is an extra gauge symmetry we can use to preserve this condition \cite{CWY}), apart from being constant at infinity along $\C$. From the form of the parametrization \eqref{sols}, the latter just correspond to the left multiplication of $g$ by $\til{\Omega}$ defined in \eqref{gxg}, with conserved current $J_{\vphi}$. This is therefore the relevant $G$-symmetry that ought to be preserved in the quantum theory.  
As we have seen in \eqref{gxg}, there are two $G$ transformations that leave \eqref{cwzw3} invariant, but the one defined by right multiplication of $g$ by $\Omega$ is in fact a redundancy arising from the form of the parametrization of $A$ by $g$   in \eqref{sols}, and ought to be fixed. 
The relevant $G$ symmetry therefore ought to 
%be the one defined by
arise from left multiplication of $g$ by $\til{\Omega}$, with conserved current $J_{\vphi}$, and this should be preserved in the quantum theory. 

In fact, rewriting \eqref{firstquantum}
as 
%his is in fact equivalent to the current algebra
\begin{equation} \label{rewritten}
\begin{aligned}
  \left[\textrm{Tr}A J_{\vphi}(\vphi,z),\textrm{Tr}B J_{\vphi}(\vphi',z')\right]=&i\dl(\vphi-\vphi')\dl(z-z')\textrm{Tr}[A,B] J_\vphi(\vphi,z)\\&-\frac{i}{\pi\hbar}\dl'(\vphi-\vphi')\dl(z-z')\textrm{Tr}AB + \ldots,
 \end{aligned}
\end{equation}
 %Integrating \eqref{firstquantum}
 and integrating over $\vphi$, $\vphi'$, $z$ and $z'$ on both sides, we find that $Q=\int d\vphi \int dz J_{\vphi}(\vphi,z)$ generates this quantum $G$ symmetry if the quantum corrections are restricted to be total derivatives, i.e., it satisfies
 \begin{equation}
 [Q^a,Q^b]=if^{ab}_cQ^c.
 \end{equation}
%\begin{equation}
%[\int dz \int d\vphi J^a_{\vphi}(\vphi,z),\int dz' \int d\vphi' J^b_{\vphi}(\vphi',z')]=if^{ab}_c\int dz \int d\vphi J^c_{\vphi}(\vphi,z).
%\end{equation}
Moreover, since $Q$ generates this symmetry, 
\begin{equation}\label{new2}
[J_{\vphi}^a(\vphi,z),g(\vphi',z')]=-iT^ag(\vphi',z')\delta(\vphi-\vphi')\delta(z-z')
\end{equation}
ought to hold up to total derivatives, since this would imply $i\widetilde{\omega}_a[Q^a,g(\vphi'z')]=\widetilde{\omega}_aT^ag(\vphi',z')$, which is the infinitesimal form of the left multiplication of $g$. 
%This can be shown to imply that \eqref{rewritten} holds up to total derivatives with respect to $\vphi$ and $z$, which vanish when it is integrated over these variables.
Note that since the corrections to \eqref{rewritten} and \eqref{new2} are restricted to be total derivatives (or repeated total derivatives) by the $G$ symmetry, these corrections must be multiplied by positive powers of $\hbar$ (which has dimensions of length) to satisfy dimensional analysis.
We expect that the explicit form of these corrections, that are higher order in $\hbar$, can be found via a Moyal quantization scheme that preserves the
%aforementioned 
$G$ symmetry.

To express \eqref{rewritten} in a more familiar form, we expand the currents in terms of their Fourier modes along $S^1$,
\begin{equation}
J_{\vphi}(\vphi,z)=\frac{1}{2\pi}\sum^{\infty}_{n=-\infty}J_{\vphi}^{n}(z)e^{in\vphi},
\end{equation}
and utilize the orthogonality of these modes, which leads us to
\begin{equation}\label{firstalgebra} 
\begin{aligned}
  \left[\textrm{Tr}A J^n_{\vphi}(z),\textrm{Tr}B J^m_{\vphi}(z')\right]=i\textrm{Tr}[A,B] J^{n+m}_\vphi(z)\dl(z-z')+\frac{2}{\hbar}(n\dl_{m+n,0})\dl(z-z')\textrm{Tr}AB+\ldots.
 \end{aligned}
\end{equation}
This algebra has the form of a Kac-Moody algebra with generators having holomorphic dependence on the Riemann surface, $\C$. Note that %a priori 
there is no quantization condition on $\hbar$ here, unlike the current algebra derived from the boundary theory of ordinary 3d Chern-Simons theory. 
%Also note that the term that contains $\hbar$, which takes the form of a central extension, arises from a central extension term that appears in the classical Poisson bracket \eqref{pbrckt}. This is analogous to the case of the standard 2d WZW model 
%with kinetic term in ``chiral'' form, i.e., using lightcone coordinates, wherein such a central extension term also appears in the classical Poisson bracket involving currents when taking one of the lightcone directions to be time, as shown in \cite{w}.
%, unlike the current algebra of the 2d WZW model.
%3d Chern-Simons theory.

To further understand this algebra, we 
shall 
%analytically continue $z$ and $\zb$ 
 %such that they take real values, and  
write $z=\eps t +i \theta$, compactify the $\theta$ direction to be valued in $[0,2\pi]$, and subsequently take $\eps\rightarrow 0$.
%we %can 
%compactify the $\zb$ direction such that it is valued in $[0,2\pi]$. 
Upon doing so, we may perform another expansion in Fourier modes, i.e., 
\begin{equation}
J^n_{\vphi}(\theta)=\frac{1}{2\pi}\sum^{\infty}_{\tilde{n}=-\infty}J_{\vphi}^{n,\tilde{n}}e^{i\tilde{n}\theta}.
\end{equation}
Then, employing the orthogonality of these modes, the resulting algebra takes the form 
\begin{equation} \label{secondalgebra}
\begin{aligned}
  \left[\textrm{Tr}A J^{n,\tilde{n}}_{\vphi},\textrm{Tr}B J^{m,\tilde{m}}_{\vphi}\right]=i\textrm{Tr}[A,B] J^{n+m,\tilde{n}+\tilde{m}}_\vphi +\frac{4\pi}{\hbar}n\dl_{m+n,0} \dl_{\tilde{m}+\tilde{n},0}\textrm{Tr}AB +\ldots.
 \end{aligned}
\end{equation}
This %is known as 
has the form of a two-toroidal Lie algebra (or a centrally-extended double loop algebra), which, in particular, arises as the current algebra of the four-dimensional WZW model studied in \cite{Donaldson, Nair, Losev, Kanno}.
Hence, the algebra \eqref{firstalgebra} that we obtained can be understood to be an `analytical continuation' of the  two-toroidal Lie algebra \eqref{secondalgebra}, with one of the two `loop directions' decompactified.  This is not surprising, considering the fact that 4d Chern-Simons theory for $C=\R\times S^1$ can be understood to be 3d Chern-Simons theory for the loop group, but with the `loop direction' complexified \cite{Witten}.

%Rough argument- Let us return to the question of higher order terms in $\hbar$ on the RHS of ... Such terms would be O(1) terms in (2.25), since to get the algebra in this form we need to divide both sides by hbar. We shall once again write $z=\eps t +i \theta$, but compactify the $\theta$ direction to be valued in $[0,2\pi R]$ now. Now, the combined $\eps \rightarrow 0$ and $R\rightarrow 0$ limit should lead us to 3d a.c. CS theory from 4d CS. So the algebra above should become a 2d affine Kac-moody algebra in the same limit. The terms that we are concerned with should therefore vanish in this limit. In other words, such terms must be multiplied by some powers of $\epsilon$ and $R$. However, $\epsilon$ can be eliminated in 4d CS theory via a rescaling (cite Witten, to check), so the algebra should not depend on this parameter. However, there can be terms multiplied by some powers of $R$, but in the $R\rightarrow \infty$ limit that we are interested in, these diverge. Such divergent terms should not occur in a physically well-defined theory such as 4d CS, which is well-defined both perturbatively as well as non-perturbatively (using Lefschetz thimbles to provide integration cycles see CY AT).

A further comparison with the four-dimensional WZW model is in order. Firstly, note that 4d Chern-Simons theory is unrenormalizable by power-counting, and so is its dual 3d ``chiral" WZW model,
% that we have been studying, 
since in both cases the loop-counting parameter, $\hbar$, has dimensions of inverse mass. %Nevertheless, i
Nevertheless, it has been shown that 4d Chern-Simons theory can be quantized in perturbation theory using BV quantization \cite{Costello2}, ensuring that it is finite. Now, the 4d WZW model is one-loop finite (and is likely to be all-loop finite) despite being unrenormalizable by power-counting \cite{Ketov}, and it also admits an infinite-dimensional current algebra. The presence of this algebra appears to be responsible for the one-loop finiteness, since  this is the only four-dimensional nonlinear sigma model that is one-loop finite apart from the sigma model on $S^7$. Thus, the infinite-dimensional algebra \eqref{firstalgebra} should likewise 
%be responsible for 
give rise to the finiteness of the 3d ``chiral" WZW model, and therefore of 4d Chern-Simons theory. It would be of interest to explore this further.
%This deserves to be explored further.
%the presence of the infinite-dimensional algebra \eqref{firstalgebra} in the boundary dual to 4d Chern-Simons theory could be the source of its finiteness, and this deserves to be explored further. 
%It remains to be directly verified that the 3d ``chiral" WZW model is finite, and in this direction, we shall 
%%do so for some one- and two-loop 
%demonstrate the finiteness of some of its free-field correlation functions in the next section.

%The presence of the infinite-dimensional current algebra \eqref{firstalgebra} in the boundary dual to 4d Chern-Simons theory appears to be responsible for this finiteness. This can be compared to the 4d WZW model \cite{Kanno}, which is also one-loop finite despite being unrenormalizable by power-counting. It remains to be directly verified that the 3d WZW model is finite, and we shall do so for to one- and two-loop amplitudes in the next section.

Finally, we shall also compare our 3d WZW model with the 2d integrable sigma models of Delduc, Lacroix, Magro and Vicedo \cite{Delduc} that admit descriptions as affine Gaudin models, and are obtained from 4d Chern-Simons theory via the approach of Costello and Yamazaki \cite{CWY3}. As an intermediate step towards deriving these sigma models in \cite{Delduc}, the action 
\begin{align} \label{action hat g}
S &=  \frac{i}{12 \pi} \int_{\Sigma \times \CP} \omega \wedge \langle \widehat{g}^{-1} d \widehat{g}, \widehat{g}^{-1} d \widehat{g} \wedge \widehat{g}^{-1} d \widehat{g} \rangle 
 + \frac{i}{4 \pi} \int_{\Sigma \times \CP} d\omega \wedge \langle \widehat{g}^{-1} d \widehat{g}, \mathcal L \rangle,
\end{align}
was obtained from 4d Chern-Simons theory (where we use the notation and conventions of \cite{Delduc}) via %a formal gauge transformation 
$A=-d\widehat{g}\widehat{g}^{-1}+\widehat{g}\mathcal{L}\widehat{g}^{-1}$, where $\mathcal{L}$ is interpreted as a Lax connection. Now, if we include a boundary for $\Sigma$ by setting $\Sigma= D$, one finds the additional term
\begin{equation}
-\frac{i}{4\pi}\int_{\del\Sigma\times \CP}\omega \wedge \langle \widehat{g}^{-1}d\widehat{g},\mathcal{L}\rangle
\end{equation}
via this procedure.
To relate to our 3d WZW model, we shall set $\omega=dz$, and we shall pick $\mathcal{L}=-\del_{\vphi}{\til{g}}{\til{g}}^{-1} d\vphi$ for a map $\til{g}\rightarrow \del \Sigma \times \CP$, where $\mathcal{L}_{\vphi}$ obeys $\del_{r}\mathcal{L}_{\vphi}=0$ and $\del_{\zb}\mathcal{L}_{\vphi}=0$ (which is reminiscent of the current conservation equation for $J_{\vphi}$). The full action can  be written as 
\begin{equation}
\begin{aligned}
S =&  \frac{i}{12 \pi} \int_{\Sigma \times \CP} dz \wedge \langle \widehat{g}^{-1} d \widehat{g}, \widehat{g}^{-1} d \widehat{g} \wedge \widehat{g}^{-1} d \widehat{g} \rangle
+\frac{i}{4\pi}\int_{\del\Sigma\times \CP}dz \wedge \langle \widehat{g}^{-1}d\widehat{g},\del_{\vphi}{\til{g}}{\til{g}}^{-1} d\vphi\rangle,
\end{aligned}
\end{equation}
where we have used the fact that $d(dz)=0$ everywhere except $z\rightarrow \infty$, which is where we take $\til{g}\rightarrow \mathds{1}$ (the same boundary condition was previously used implicitly for the field $g$). Then, taking $\widehat{g}={\bar{g}}^{-1}$, and restricting $\bar{g}$ to be equal to $\til{g}$ along $\del \Sigma$, we find 
\begin{equation}
\begin{aligned}
S =&  -\frac{i}{12 \pi} \int_{\Sigma \times \CP} dz \wedge \langle d \bar{g}\bar{g}^{-1}, d \bar{g}\bar{g}^{-1} \wedge  d \bar{g}\bar{g}^{-1} \rangle -\frac{i}{4\pi}\int_{\del\Sigma\times \CP} d\vphi \wedge dz \wedge d\zb \langle \del_{\zb}\til{g}\til{g}^{-1},\del_{\vphi}{\til{g}}{\til{g}}^{-1} \rangle,
\end{aligned}
\end{equation}
which is a 3d ``chiral" WZW action, where the equation of motion $\del_{\zb}(\del_{\vphi}{\til{g}}{\til{g}}^{-1})=0$ has been implemented.\footnote{Note that using our earlier conventions for the 4d Chern-Simons action, we can obtain the 3d WZW action \eqref{cwzw3} in the same manner, multiplied by a minus sign that can be absorbed into the definition of $\hbar$.} Thus, an analogous derivation starting with 4d Chern-Simons theory on $D\times \CP$ with a more general choice of $\omega$ ought to lead to a generalization of the 3d WZW model, which we expect to be equivalent, via a generalization of the approach of \cite{CWY3,Delduc}, to a 2d integrable sigma model with a 1d boundary action. Based on the results of \cite{Delduc,Vicedo}, this integrable sigma model is likely to be described by some type of affine Gaudin model.
\section{Wilson Lines and Boundary Local Operators}
We shall now describe Wilson lines in 4d Chern-Simons theory in terms of local operators of its boundary dual.
%the boundary theory. 
This is possible due to the flatness condition that restricts the gauge field components on $D$ to be pure gauge configurations, as shown in \eqref{sols}.
To see this, note that a Wilson line along a curve $\mathcal{C}$, starting at $t_i$ and ending at $t_f$, and in a representation $R$, satisfies
\begin{equation}
\mathcal{P} e^{\int_{\mathcal{C}} \cA}=g_R^{-1}(t_f)\mathcal{P} e^{\int_{\mathcal{C}} \cA'}g_R(t_i)
\end{equation}
where $\cA=g\cA'g^{-1}-dgg^{-1}$. Setting $\cA'=0$, we find that 
\begin{equation}\label{hol}
\mathcal{P} e^{\int_{\mathcal{C}} (-dgg^{-1})}=g^{-1}_R(t_f)g_R(t_i).
\end{equation}
If $t_f$ and $t_i$ are points on $\del D=S^1$, this implies that a bulk Wilson line operator can be completely described in terms of a pair of local boundary operators. In fact, such boundary-anchored Wilson lines are automatically gauge invariant, since all gauge transformations are trivial at the boundary.

This suggests that correlation functions of Wilson line operators in 4d Chern-Simons theory can be captured by correlation functions of local operators in the 3d ``chiral" WZW model. This includes correlators of crossed Wilson lines which compute the $R$-matrices of integrable lattice models. We shall now attempt to verify this.

\begin{figure}[!h]
\centering
\begin{tikzpicture}[scale=3]
% \draw[step=.5cm, gray, very thin] (-1.2,-1.2) grid (1.2,1.2); 
 %\filldraw[fill=green!20,draw=green!50!black] (0,0) -- ({(cos(60))*3 mm} ,{(sin(60))*3 mm} ) arc (60:90:3mm) -- cycle node[above,xshift=3mm,yshift=9mm]{$\delta$}; 
 \draw[gray] (-1.0,0) -- (1.0,0) coordinate (x axis);
 \draw[gray] (0,-1.0) -- (0,1.0) coordinate (y axis);
 \draw (0,0) circle (1cm);
 %\draw[very thick,red] (30:1cm) -- node[left,fill=white] {$\sin \alpha$} (30:1cm |- x axis);
 %\draw[very thick,blue] (30:1cm |- x axis) -- node[below=2pt,fill=white] {$\cos \alpha$} (0,0);
 \draw [very thick,blue] (0,0) -- (90:1cm);
 \draw [very thick,blue] (0,0) -- (270:1cm);
  \draw [very thick,blue] (0,0) -- (0:1cm);
 \draw [very thick,blue] (0,0) -- (180:1cm);
 %\foreach \x/\xtext in {-1, -0.5/-\frac{1}{2}, 1} 
%   \draw (\x cm,1pt) -- (\x cm,-1pt) node[anchor=north,fill=white] {$\xtext$};
 %\foreach \y/\ytext in {-1, -0.5/-\frac{1}{2}, 0.5/\frac{1}{2}, 1} 
  % \draw (1pt,\y cm) -- (-1pt,\y cm) node[anchor=east,fill=white] {$\ytext$};
  \node[label=right:{$g^{-1}_{R_1}(0,z_1,\zb_1)$},shape=circle,fill=black, scale=0.5] at (1,0) {};
   \node[label=left:{$g_{R_1}(\pi,z_1,\zb_1)$},shape=circle,fill=red, scale=0.5] at (-1,0) {};
    \node[label=above:{$g^{-1}_{R_2}(\frac{\pi}{2},z_2,\zb_2)$},shape=circle,fill=black, scale=0.5] at ({(cos(90))},{(sin(90))}) {};
   \node[label=below:{$g_{R_2}(\frac{3\pi}{2},z_2,\zb_2)$},shape=circle,fill=red, scale=0.5] at (-{(cos(90))},-{(sin(90))}) {};
 \end{tikzpicture}
 \caption{Perpendicular Wilson lines on $D$.}
 \label{perp}
\end{figure}
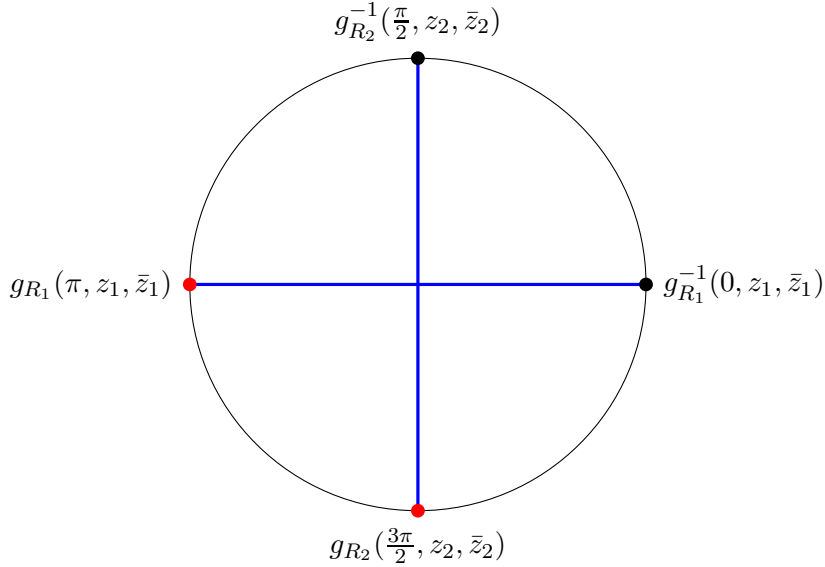

In the bulk computation by Costello, Witten, and Yamazaki \cite{CWY}, the order $\hbar$ contribution to the $R$-matrix was found by performing perturbation theory around the trivial field configuration, $\cA=0$, and computing free-field propagators between Wilson lines. In the same vein, we shall consider perturbation theory around the field configuration $g=\mathds{1}$, and we shall use a free-field propagator to compute the relevant correlation function involving operators appearing on the right hand side of \eqref{hol}, i.e.,
 \begin{equation}\label{eqeq}
 \begin{aligned}
 &\langle \mathcal{P} e^{\int_{\varphi=\pi}^{\varphi=0} \cA_{R_1}(z_1,\zb_1)}\otimes \mathcal{P} e^{\int^{\varphi=\pi/2}_{\varphi=3\pi/2}\cA_{R_2}(z_2,\zb_2)}\rangle \\=& \langle g_{R_1}^{-1}(0,z_1,\zb_1)g_{R_1}(\pi,z_1,\zb_1)\otimes g_{R_2}^{-1}({\pi}/{2},z_2,\zb_2)g_{R_2}({3\pi}/{2},z_2,\zb_2) \rangle.
 \end{aligned}
   \end{equation}
  The relevant operators are depicted in Figure \ref{perp}.
   %via the following diagram.
   
%\usetikzlibrary{arrows}   
 %  \newcommand{\midarrow}{\tikz \draw[-triangle 90] (0,0) -- +(.1,0);}
%\begin{tikzpicture}
    %% equidistant points and arc
    %\foreach \x [count=\p] in {0,...,11} {
     %   \node[shape=circle,fill=black, scale=0.5] (\p) at (-\x*90:2) {};};
    %%\foreach \x [count=\p] in {0,...,5} {
     %   %\draw (-\x*60:2.4) node {\p};
      %  %\draw (-30-\x*60:2.4) node {$\bar{\p}$};}; 
   % \draw (1) arc (0:360:2);
    %   \draw[black] (1) -- (7) ; %%node{\midarrow} 
 %   \draw[black] (10) -- (4) ;%node {\midarrow} 
  %  % axes
   % %\draw [dotted, gray] (-2.6,0) -- (2.6,0);
    %%\draw [dotted, gray] (0,-2.15) -- (0,2.15);
%\end{tikzpicture}
%\\

Now, %since we are considering perturbations around $g=\mathds{1}$, 
we may expand the field, $g$, as $$g=e^{\phi_aT^a}=\mathds{1}+\phi_aT^a +\ldots$$
% and neglect higher order terms represented by the ellipsis. Thus, in this approximation, t
The free part of the 3d WZW action is then
\begin{equation}
\frac{1}{2\pi\hbar}\int_{S^1\times \C} d\varphi \wedge dz \wedge  d\zb\textrm{Tr}\big(\del_{\varphi}gg^{-1}\del_{\zb}gg^{-1})=-\frac{1}{2\pi\hbar}\int_{S^1\times \C}d\varphi \wedge dz \wedge d\zb ~~\phi^a\del_{\varphi}\del_{\zb}\phi_a + \ldots,
\end{equation}
where we have performed integration by parts after expanding the field $g$. 

Next, from the %leading order term of the
 free action, one may construct the generating functional
\begin{equation}\label{gf}
\begin{aligned}
Z_0[J]&=\frac{\int D\phi e^{\frac{i}{2\pi\hbar}\int_{S^1\times \C}d\varphi \wedge dz \wedge  d\zb(-\phi^a\del_{\vphi}\del_{\zb}\phi_a +2\pi i\hbar J_a\phi^a)}}{\int D\phi e^{\frac{i}{2\pi\hbar}\int_{S^1\times \C}d\varphi \wedge dz \wedge  d\zb(-\phi^a\del_{\vphi}\del_{\zb}\phi_a )}}\\
&=\textrm{exp}\bigg(-\frac{2\pi i\hbar}{4}\int d^3x \int d^3y J_a(x)\Delta^{ab}(x-y) J_b(y)\bigg),
\end{aligned}
\end{equation}
where $x=(\varphi, z,\zb)$, $y=(\varphi', z',\zb')$, and $\Delta^{ab}$ is the propagator which obeys
\begin{equation}\label{green}
\del_{\varphi}\del_{\zb}\Delta^{ab}(x)=\delta^{ab}\delta^3(x),
\end{equation}
and is given explicitly by 
\begin{equation}\label{propa}
\Delta^{ab}(x)=\delta^{ab}\frac{1}{2\pi i}\frac{1}{z}\til{\Delta}_{\vphi}
\end{equation}
%where 
%$$\delta(z)= \frac{1}{2\pi i}\frac{1}{z},$$
%which behaves as a delta function via Cauchy's integral formula, 
where 
\begin{equation}\label{mult}
\widetilde{\Delta}_{\vphi}=\frac{1}{2\pi}\bigg(\sum_{k=1}^{\infty}\frac{e^{ik\vphi}}{ik}+\vphi+\sum_{k=-\infty}^{-1}\frac{e^{ik\vphi}}{ik}\bigg),
\end{equation}
which satisfies $\del_{\vphi}\widetilde{\Delta}_{\vphi}=\frac{1}{2\pi}\big(\sum_{k=-\infty}^{\infty}{e^{ik\vphi}}\big)=\delta(\vphi)$. 
%In particular, 
%\begin{equation}
%\til{\Delta}_{\frac{\pi}{2}}=\frac{1}{2\pi}\frac{\pi}{2}+\frac{1}{\pi}\bigg(1-\frac{1}{3}+\frac{1}{5}-\frac{1}{7}\ldots\bigg)=\frac{1}{2},
%\end{equation}
%and likewise $\til{\Delta}_{-\frac{\pi}{2}}=-\frac{1}{2}$. 
Note that %we use different definitions for $\delta(z)$ and $\delta(\zb)$ in order to ensure that
\eqref{green} 
holds since 
\begin{equation}
\del_{\zb}\frac{1}{2\pi i}\frac{1}{z}=\delta^2(z,\zb).
\end{equation}
Moreover, the propagator satisfies  $\Delta^{ab}(x-y)=\Delta^{ba}(y-x)$.
The two point function can be found from \eqref{gf} to be 
\begin{equation}
\langle \phi^a({x})\phi^b({y})\rangle=-\pi i\hbar\Delta^{ab}({x}-{y}).
\end{equation}

Before we proceed, we first note that the propagator \eqref{propa} has no dependence on the $\zb$ coordinate. This can be understood from the fact that there is a gauge redundancy in the 3d ``chiral" WZW model, namely the invariance of the parametrizations \eqref{sols} under transformations generated by $\Omega$ in \eqref{gxg}. 
%Fixing of this redundancy can be achieved 
This redundancy can be fixed by setting $B(z,\zb)=\mathds{1}$ in \eqref{eomsol}. As a result, the operator $g$ has no $\zb$-dependence, and therefore correlation functions involving it should not have $\zb$-dependence either.

Another pertinent point to note is that the propagator is a multi-valued function, as it includes the expression on the RHS of\eqref{mult} as a factor. Hence, to obtain a single-valued propagator, we ought to define it with a branch cut. We shall pick the branch cut to be from $r=0$ to ($r=R$, $\vphi=\pi$), where $R$ is the radius of $D$. This effectively restricts $\vphi$ in \eqref{mult} to take values in $(-{\pi},{\pi})$. In this manner, we obtain a well-defined, single-valued propagator.

%we note that there is a gauge redundancy to be fixed, namely, the dependence of the fields on $\zb$. This can be observed from \eqref{gxg}. Here $\til{\Omega}$ generates a global symmetry, since it does not go to $\mathds{1}$ at infinity, while ${\Omega}$ generates a local symmetry. Since $\zb$-dependence is only possible in transformations generated by $\Omega$, this implies that $\zb$-dependence is a gauge redundancy that can be eliminated via an appropriate choice of $\Omega$. Thus, in what follows, we shall set $\zb=0$ in describing physical boundary modes, and as a result we shall drop the (divergent) factor $\frac{\int dp}{2\pi i}\frac{e^{ip\zb}}{p}$ from \eqref{propa} when computing correlators.

Now, to compute the RHS of \eqref{eqeq}, we shall expand each operator to linear order in $\phi$
\begin{equation}\label{linord}
\begin{aligned}
&\langle g^{-1}_{R_1}(0,z_1)g_{R_1}(\pi,z_1)\otimes g^{-1}_{R_2}({\pi}/{2},z_2)g_{R_2}({3\pi}/{2},z_2) \rangle \\=& \langle(\mathds{1}-\phi_a(0,z_1)T^a_{R_1})(\mathds{1}+\phi_b(\pi,z_1)T^b_{R_1})\otimes(\mathds{1}-\phi_c(\pi/2,z_2)T^c_{R_2})(\mathds{1}+\phi_d(3\pi/2,z_2)T^d_{R_2})\rangle + \ldots
\end{aligned}
\end{equation}
We then only keep terms of quadratic or lower order in the fields, while taking self-contractions (i.e., correlators of operators with the same value of $z$) to be zero via regularization. Also note that 1-point functions can be shown to be zero using \eqref{gf}. Hence, we find
\begin{equation}\label{res1}
\begin{aligned}
&\mathds{1}+\langle \phi_a(0,z_1)\phi_c(\pi/2,z_2)\rangle T_{R_1}^a\otimes T_{R_2}^c-\langle \phi_a(\pi,z_1)\phi_c(\pi/2,z_2) \rangle T_{R_1}^a\otimes T_{R_2}^c \\-&\langle \phi_a(2\pi,z_1)\phi_c(3\pi/2,z_2) \rangle T_{R_1}^a\otimes T_{R_2}^c+\langle \phi_a(\pi,z_1)\phi_c(3\pi/2,z_2)\rangle T_{R_1}^a\otimes T_{R_2}^c + O(\hbar^2)\\
=&\mathds{1}-\frac{\hbar}{2}\delta_{ac} \frac{1}{z_1-z_2}\til{\Delta}_{-\frac{\pi}{2}}T_{R_1}^a\otimes T_{R_2}^c+\frac{\hbar}{2}\delta_{ac}\frac{1}{z_1-z_2}\til{\Delta}_{\frac{\pi}{2}}T_{R_1}^a\otimes T_{R_2}^c\\&+\frac{\hbar}{2}\delta_{ac}\frac{1}{z_1-z_2}\til{\Delta}_{\frac{\pi}{2}}T_{R_1}^a\otimes T_{R_2}^c-\frac{\hbar}{2}\delta_{ac}\frac{1}{z_1-z_2}\til{\Delta}_{-\frac{\pi}{2}}T_{R_1}^a\otimes T_{R_2}^c + O(\hbar^2)\\
=&\mathds{1}+\frac{\hbar}{z_1-z_2}(\til{\Delta}_{\frac{\pi}{2}}-\til{\Delta}_{-{\frac{\pi}{2}}})T_{R_1}^a\otimes T_{ R_2a} + O(\hbar^2)\\
=&\mathds{1}+\frac{\hbar}{z_1-z_2}T_{R_1}^a\otimes T_{ R_2 a} + O(\hbar^2).
\end{aligned}
\end{equation}
Here, we have used the fact that  
\begin{equation}
\til{\Delta}_{\frac{\pi}{2}}=\frac{1}{2\pi}\frac{\pi}{2}+\frac{1}{\pi}\bigg(1-\frac{1}{3}+\frac{1}{5}-\frac{1}{7}\ldots\bigg)=\frac{1}{2},
\end{equation}
and likewise $\til{\Delta}_{-\frac{\pi}{2}}=-\frac{1}{2}$. We have thus obtained, from our 3d ``chiral" WZW model, the \textit{exact} order $\hbar$ correlation function for a pair of perpendicular Wilson lines that Costello, Witten and Yamazaki \cite{CWY} computed via the bulk 4d Chern-Simons path integral. 
%Note that our convention for the 4d CS action differs from that of Costello-Witten-Yamazaki by an overall factor of $\frac{1}{2\pi i}$. Hence, to agree with their convention, we rescale $\hbar$ by $2\pi i$, which gives us exactly the order $\hbar$ result they obtain via a bulk path integral computation.

%Once the order $\hbar$ contribution to the $R$-matrix is obtained, as we have done above, the full expression for the $R$-matrix follows from general theorems (up to a change of variables) \cite{Drinfeld,Chari}.
\subsection{Non-perpendicular Wilson Lines}
\begin{figure}[!h]
\centering
\begin{tikzpicture}[scale=3]
% \draw[step=.5cm, gray, very thin] (-1.2,-1.2) grid (1.2,1.2); 
 \filldraw[fill=green!20,draw=green!50!black] (0,0) -- ({(cos(60))*3 mm} ,{(sin(60))*3 mm} ) arc (60:90:3mm) -- cycle node[above,xshift=3mm,yshift=9mm]{$\delta$}; 
 \draw[gray] (-1.0,0) -- (1.0,0) coordinate (x axis);
 \draw[gray] (0,-1.0) -- (0,1.0) coordinate (y axis);
 \draw (0,0) circle (1cm);
 %\draw[very thick,red] (30:1cm) -- node[left,fill=white] {$\sin \alpha$} (30:1cm |- x axis);
 %\draw[very thick,blue] (30:1cm |- x axis) -- node[below=2pt,fill=white] {$\cos \alpha$} (0,0);
 \draw [very thick,blue] (0,0) -- (60:1cm);
 \draw [very thick,blue] (0,0) -- (240:1cm);
  \draw [very thick,blue] (0,0) -- (0:1cm);
 \draw [very thick,blue] (0,0) -- (180:1cm);
 %\foreach \x/\xtext in {-1, -0.5/-\frac{1}{2}, 1} 
%   \draw (\x cm,1pt) -- (\x cm,-1pt) node[anchor=north,fill=white] {$\xtext$};
 %\foreach \y/\ytext in {-1, -0.5/-\frac{1}{2}, 0.5/\frac{1}{2}, 1} 
  % \draw (1pt,\y cm) -- (-1pt,\y cm) node[anchor=east,fill=white] {$\ytext$};
  \node[label=right:{$g^{-1}_{R_1}(0,z_1)$},shape=circle,fill=black, scale=0.5] at (1,0) {};
   \node[label=left:{$g_{R_1}(\pi,z_1)$},shape=circle,fill=red, scale=0.5] at (-1,0) {};
    \node[label={[label distance=0.5mm]60:$g_{R_2}^{-1}(\frac{\pi}{2}-\delta,z_2)$},shape=circle,fill=black, scale=0.5] at ({(cos(60))},{(sin(60))}) {};
   \node[label={[label distance=0.5mm]240:$g_{R_2}(\frac{3\pi}{2}-\delta,z_2)$},shape=circle,fill=red, scale=0.5] at (-{(cos(60))},-{(sin(60))}) {};
 \end{tikzpicture}
\caption{Non-perpendicular Wilson lines on $D$.}
\label{nonperp}
 \end{figure}
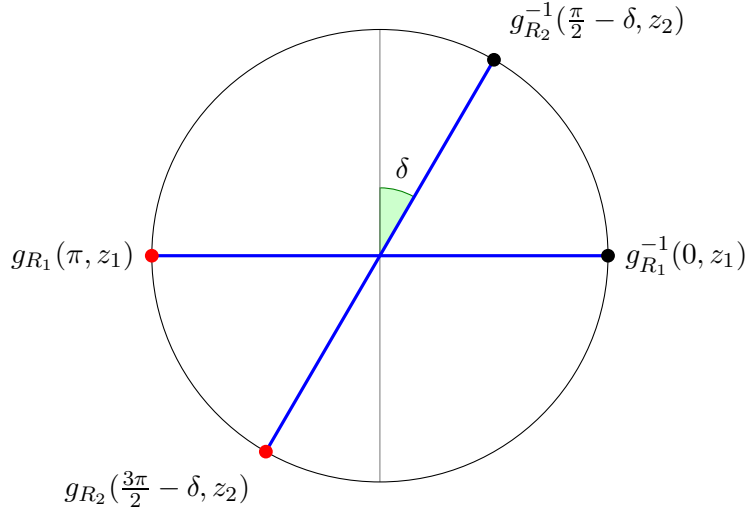

We can generalize the calculation above to the case of non-perpendicular Wilson lines. As an example, we shall start with two perpendicular Wilson lines, and rotate the vertical Wilson line clockwise by an angle, $\delta$, as shown in Figure \ref{nonperp}.

The four-point function we should compute is 
\begin{equation}
\begin{aligned}
&\langle g_{R_1}^{-1}(0,z_1)g_{R_1}(\pi,z_1)\otimes g_{R_2}^{-1}({\pi}/{2}-\delta,z_2)g_{R_2}({3\pi}/{2}-\delta,z_2) \rangle.
\end{aligned}
\end{equation}
Expanding each operator to linear order in $\phi$ as in \eqref{linord}, we find 
\begin{equation}\label{callcc2}
\begin{aligned}
&\mathds{1}+\langle \phi_a(0,z_1)\phi_c(\pi/2-\delta,z_2)\rangle T_{R_1}^a\otimes T_{R_2}^c-\langle \phi_a(\pi,z_1)\phi_c(\pi/2-\delta,z_2)\rangle T_{R_1}^a\otimes T_{R_2}^c\\-&\langle \phi_a(2\pi,z_1)\phi_c(3\pi/2-\delta,z_2)\rangle T_{R_1}^a\otimes T_{R_2}^c+\langle \phi_a(\pi,z_1)\phi_c(3\pi/2-\delta,z_2)\rangle T_{R_1}^a\otimes T_{R_2}^c +O(\hbar^2)\\
=&\mathds{1}-\frac{\hbar}{2}\delta_{ac} \frac{1}{z_1-z_2}\til{\Delta}_{-\frac{\pi}{2}+\delta}T_{R_1}^a\otimes T_{R_2}^c+\frac{\hbar}{2}\delta_{ac} \frac{1}{z_1-z_2}\til{\Delta}_{\frac{\pi}{2}+\delta}T_{R_1}^a\otimes T_{R_2}^c\\&+\frac{\hbar}{2}\delta_{ac} \frac{1}{z_1-z_2}\til{\Delta}_{\frac{\pi}{2}+\delta}T_{R_1}^a\otimes T_{R_2}^c-\frac{\hbar}{2}\delta_{ac} \frac{1}{z_1-z_2}\til{\Delta}_{-\frac{\pi}{2}+\delta}T_{R_1}^a\otimes T_{R_2}^c +O(\hbar^2)\\
=&\mathds{1}+ \frac{\hbar}{z_1-z_2}(\til{\Delta}_{\frac{\pi}{2}+\delta}-\til{\Delta}_{-\frac{\pi}{2}+\delta})T_{R_1}^a\otimes T_{R_2a}+O(\hbar^2)
\end{aligned}
\end{equation}
Now, note that \eqref{mult} can be rewritten as 
\begin{equation}
\til{\Delta}_{\vphi}=\frac{\vphi}{2\pi}+\frac{1}{\pi}\sum^{\infty}_{k=1}\frac{\textrm{sin}(k\vphi)}{k}.
\end{equation}
This implies that 
\begin{equation}\label{ff}
\til{\Delta}_{\frac{\pi}{2}+\delta}=\frac{\frac{\pi}{2}+\delta}{2\pi}+\frac{1}{\pi}\sum^{\infty}_{k=1}\frac{\textrm{sin}(k\frac{\pi}{2})\textrm{cos}(k\delta)+\textrm{cos}(k\frac{\pi}{2})\textrm{sin}(k\delta)}{k},
\end{equation}
and
\begin{equation}
\til{\Delta}_{-\frac{\pi}{2}+\delta}=\frac{-\frac{\pi}{2}+\delta}{2\pi}-\frac{1}{\pi}\sum^{\infty}_{k=1}\frac{\textrm{sin}(k\frac{\pi}{2})\textrm{cos}(k\delta)-\textrm{cos}(k\frac{\pi}{2})\textrm{sin}(k\delta)}{k}.
\end{equation}
We find that the sums over $k$ above can be separated into two types of sums, each having the form of a Fourier series, namely, the Fourier series for a square wave, 
\begin{equation}\label{gd}
\begin{aligned}
\sum^{\infty}_{k=1}\frac{\textrm{sin}(\frac{k\pi}{2})\textrm{cos}(kx)}{k}&=\frac{\pi}{4}\textrm{sign}(\textrm{cos}(x)),
\end{aligned}
\end{equation} 
and the Fourier series for a sawtooth wave, 
\begin{equation}\label{gd3}
\begin{aligned}
\sum^{\infty}_{k=1}\frac{\textrm{cos}(\frac{k\pi}{2})\textrm{sin}(kx)}{k}&=\frac{-x}{2}+\frac{l\pi}{2}\textrm{,      }~~\pi(l-\frac{1}{2})<x<\pi(l+\frac{1}{2}), l\in \Z, 
\end{aligned}
\end{equation} 
for $x\in \R$.
However, single-valuedness of the propagators involved in the computation \eqref{callcc2} requires that $-\frac{\pi}{2}<\delta <\frac{\pi}{2}$, implying 
\begin{equation}\label{gd2}
\begin{aligned}
\sum^{\infty}_{k=1}\frac{\textrm{sin}(\frac{k\pi}{2})\textrm{cos}(k\dl)}{k}&=\frac{\pi}{4},
\end{aligned}
\end{equation} 
and 
\begin{equation}\label{gd4}
\begin{aligned}
\sum^{\infty}_{k=1}\frac{\textrm{cos}(\frac{k\pi}{2})\textrm{sin}(k\dl)}{k}=-\frac{\dl}{2}.
\end{aligned}
\end{equation}
From here we find that
\begin{equation}\label{idn}
\begin{aligned}
\til{\Delta}_{\frac{\pi}{2}+\delta}&=\frac{1}{2}\\
\til{\Delta}_{-\frac{\pi}{2}+\delta}&=-\frac{1}{2}
\end{aligned}
\end{equation} 
for $-\frac{\pi}{2}<\delta <\frac{\pi}{2}$. As a result, \eqref{callcc2} is in fact independent of the angle $\delta$, and agrees precisely with the result we found for perpendicular Wilson lines. Hence, we once again find agreement with the results of Costello, Witten and Yamazaki \cite{CWY}.
\newline
 \newline
 	\noindent{\textit{Arbitrarily Crossed Wilson Lines}}
 	
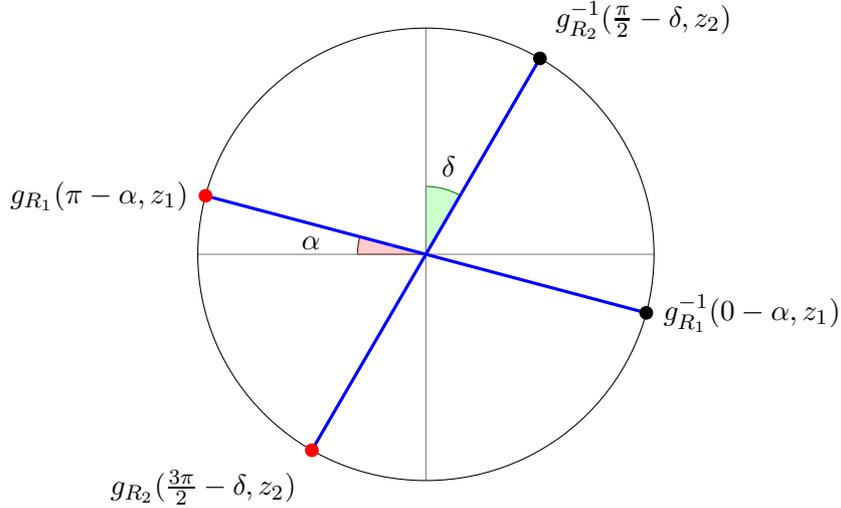
\begin{figure}[!h]
\centering
\begin{tikzpicture}[scale=3]
% \draw[step=.5cm, gray, very thin] (-1.2,-1.2) grid (1.2,1.2); 
 \filldraw[fill=green!20,draw=green!50!black] (0,0) -- ({(cos(60))*3 mm} ,{(sin(60))*3 mm} ) arc (60:90:3mm) -- cycle node[above,xshift=3mm,yshift=9mm]{$\delta$}; 
  \filldraw[fill=red!20,draw=red!50!black] (0,0) -- ({-3 mm} ,{0 mm} ) arc (180:165:3mm) -- cycle node[left,xshift=-12.5mm,yshift=1.5mm]{$\alpha$}; 
 \draw[gray] (-1.0,0) -- (1.0,0) coordinate (x axis);
 \draw[gray] (0,-1.0) -- (0,1.0) coordinate (y axis);
 \draw (0,0) circle (1cm);
 %\draw[very thick,red] (30:1cm) -- node[left,fill=white] {$\sin \alpha$} (30:1cm |- x axis);
 %\draw[very thick,blue] (30:1cm |- x axis) -- node[below=2pt,fill=white] {$\cos \alpha$} (0,0);
 \draw [very thick,blue] (0,0) -- (60:1cm);
 \draw [very thick,blue] (0,0) -- (240:1cm);
  \draw [very thick,blue] (0,0) -- (-15:1cm);
 \draw [very thick,blue] (0,0) -- (165:1cm);
 %\foreach \x/\xtext in {-1, -0.5/-\frac{1}{2}, 1} 
%   \draw (\x cm,1pt) -- (\x cm,-1pt) node[anchor=north,fill=white] {$\xtext$};
 %\foreach \y/\ytext in {-1, -0.5/-\frac{1}{2}, 0.5/\frac{1}{2}, 1} 
  % \draw (1pt,\y cm) -- (-1pt,\y cm) node[anchor=east,fill=white] {$\ytext$};
  \node[label=right:{$g^{-1}_{R_1}(0-\alpha,z_1)$},shape=circle,fill=black, scale=0.5] at ({(cos(-15))},{(sin(-15))}) {};
   \node[label=left:{$g_{R_1}(\pi-\alpha,z_1)$},shape=circle,fill=red, scale=0.5] at (-{(cos(15))},{(sin(15))}) {};
    \node[label={[label distance=0.5mm]60:$g^{-1}_{R_2}(\frac{\pi}{2}-\delta,z_2)$},shape=circle,fill=black, scale=0.5] at ({(cos(60))},{(sin(60))}) {};
   \node[label={[label distance=0.5mm]240:$g_{R_2}(\frac{3\pi}{2}-\delta,z_2)$},shape=circle,fill=red, scale=0.5] at (-{(cos(60))},-{(sin(60))}) {};
 \end{tikzpicture}
\caption{Crossed Wilson lines, both rotated from perpendicularity.}
\label{twoang}
 \end{figure}

%\section 
%\subsection{Arbitrarily Crossed Wilson Lines}
We can generalize the preceding calculations further to more general configurations of crossed Wilson lines, for which we expect to obtain the same result as \eqref{res1} due to the topological invariance of 4d Chern-Simons along $D$.
% inserted arbitrarily on the disk. 
For instance, we can consider \textit{both} Wilson lines rotated from perpendicularity, as shown in Figure \ref{twoang}.
The corresponding four-point function is unaffected by the additional rotation, i.e., we find
\begin{equation}
\begin{aligned}
&\langle g_{R_1}^{-1}(0-\alpha,z_1)g_{R_1}(\pi-\alpha,z_1)\otimes g_{R_2}^{-1}({\pi}/{2}-\delta,z_2)g_{R_2}({3\pi}/{2}-\delta,z_2) \rangle\\=&\mathds{1}+ \frac{\hbar}{z_1-z_2}(\til{\Delta}_{\frac{\pi}{2}-\alpha+\dl}-\til{\Delta}_{-{\frac{\pi}{2}-\alpha+\dl}})T_{R_1}^a\otimes T_{R_2a} +O(\hbar^2)
\\=&\mathds{1}+ \frac{\hbar}{z_1-z_2}T_{R_1}^a\otimes T_{R_2a}+O(\hbar^2),
\end{aligned}
\end{equation}
(where $-\frac{\pi}{2}<-\alpha+\dl<\frac{\pi}{2}$ to ensure single-valued propagators) with the use of the identity \eqref{idn}.

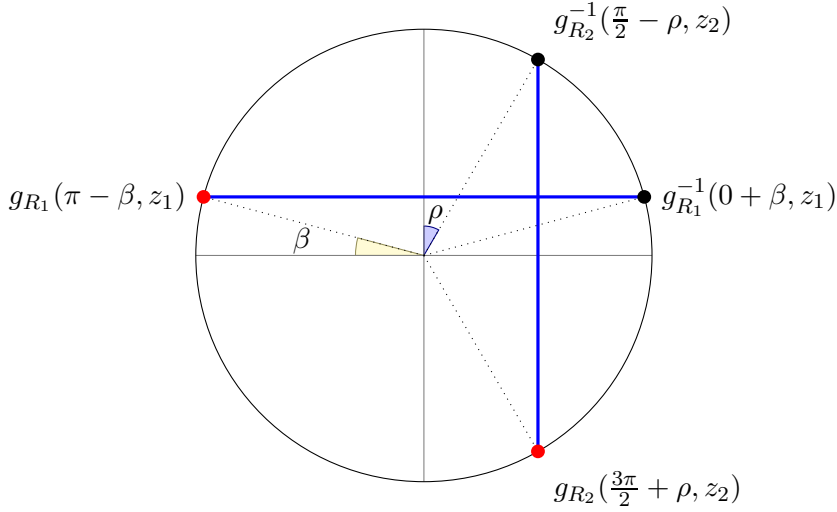
\begin{figure}[!h]
\centering
\begin{tikzpicture}[scale=3]
% \draw[step=.5cm, gray, very thin] (-1.2,-1.2) grid (1.2,1.2); 
 \filldraw[fill=blue!20,draw=blue!50!black] (0,0) -- ({(cos(60))*1.3 mm} ,{(sin(60))*1.3 mm} ) arc (60:90:1.3mm) -- cycle node[above,xshift=1.5mm,yshift=2.9mm]{$\rho$}; 
  \filldraw[fill=yellow!20,draw=yellow!50!black] (0,0) -- ({-3 mm} ,{0 mm} ) arc (180:165:3mm) -- cycle node[left,xshift=-13.5mm,yshift=2mm]{$\beta$}; 
 \draw[gray] (-1.0,0) -- (1.0,0) coordinate (x axis);
 \draw[gray] (0,-1.0) -- (0,1.0) coordinate (y axis);
 \draw (0,0) circle (1cm);
 %\draw[very thick,red] (30:1cm) -- node[left,fill=white] {$\sin \alpha$} (30:1cm |- x axis);
 %\draw[very thick,blue] (30:1cm |- x axis) -- node[below=2pt,fill=white] {$\cos \alpha$} (0,0);
 \draw [dotted] (0,0) -- (60:1cm);
 \draw [dotted] (0,0) -- (300:1cm);
\draw [very thick,blue] (60:1cm) -- (300:1cm); 
  \draw [dotted] (0,0) -- (15:1cm);
 \draw [dotted] (0,0) -- (165:1cm);
  \draw [very thick,blue] (15:1cm) -- (165:1cm);
 %\foreach \x/\xtext in {-1, -0.5/-\frac{1}{2}, 1} 
%   \draw (\x cm,1pt) -- (\x cm,-1pt) node[anchor=north,fill=white] {$\xtext$};
 %\foreach \y/\ytext in {-1, -0.5/-\frac{1}{2}, 0.5/\frac{1}{2}, 1} 
  % \draw (1pt,\y cm) -- (-1pt,\y cm) node[anchor=east,fill=white] {$\ytext$};
  \node[label=right:{$g_{R_1}^{-1}(0+\beta,z_1)$},shape=circle,fill=black, scale=0.5] at ({(cos(-15))},{(sin(15))}) {};
   \node[label=left:{$g_{R_1}(\pi-\beta,z_1)$},shape=circle,fill=red, scale=0.5] at (-{(cos(15))},{(sin(15))}) {};
    \node[label={[label distance=0.5mm]60:$g_{R_2}^{-1}(\frac{\pi}{2}-\rho,z_2)$},shape=circle,fill=black, scale=0.5] at ({(cos(60))},{(sin(60))}) {};
   \node[label={[label distance=0.5mm]300:$g_{R_2}(\frac{3\pi}{2}+\rho,z_2)$},shape=circle,fill=red, scale=0.5] at ({(cos(60))},-{(sin(60))}) {};
 \end{tikzpicture}
\caption{Perpendicular Wilson lines crossed away from the origin.}
\label{offc}
 \end{figure}

A different generalization is that of perpendicular Wilson lines crossing at a point that is not the origin, $r=0$, as shown in Figure \ref{offc}. The four-point function in this case is also independent of the angles $\beta$ and $\rho$ shown in the figure, i.e., we have
\begin{equation}
\begin{aligned}
&\langle g_{R_1}^{-1}(0+\beta,z_1)g_{R_1}(\pi-\beta,z_1)\otimes g_{R_2}^{-1}({\pi}/{2}-\rho,z_2)g_{R_2}({3\pi}/{2}+\rho,z_2) \rangle\\=&\mathds{1}+ \frac{\hbar}{z_1-z_2}\frac{1}{2}(\til{\Delta}_{\frac{\pi}{2}+\beta-\rho}-\til{\Delta}_{-{\frac{\pi}{2}+\beta+\rho}}+\til{\Delta}_{\frac{\pi}{2}-\beta+\rho}-\til{\Delta}_{-{\frac{\pi}{2}-\beta-\rho}})T_{R_1}^a\otimes T_{R_2a}+O(\hbar^2)
\\=&\mathds{1}+ \frac{\hbar}{z_1-z_2}T_{R_1}^a\otimes T_{ R_2a}+O(\hbar^2),
\end{aligned}
\end{equation}
using the identity \eqref{idn}, where $-\frac{\pi}{2}<\beta+\rho<\frac{\pi}{2}$ and $-\frac{\pi}{2}<\beta-\rho<\frac{\pi}{2}$ to ensure single-valuedness of propagators. Note that the allowed ranges of $\beta+\rho$ and $\beta-\rho$  mean that the result is valid only when the Wilson lines are crossed.
 
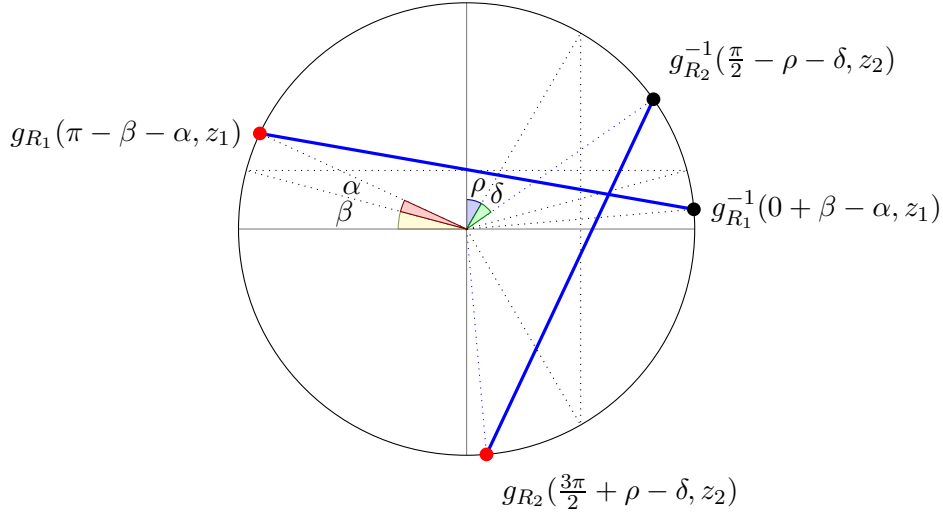
\begin{figure}[!h]
\centering
\begin{tikzpicture}[scale=3]
% \draw[step=.5cm, gray, very thin] (-1.2,-1.2) grid (1.2,1.2); 
 \filldraw[fill=blue!20,draw=blue!50!black] (0,0) -- ({(cos(60))*1.3 mm} ,{(sin(60))*1.3 mm} ) arc (60:90:1.3mm) -- cycle node[above,xshift=1.5mm,yshift=2.9mm]{$\rho$}; 
  \filldraw[fill=yellow!20,draw=yellow!50!black] (0,0) -- ({-3 mm} ,{0 mm} ) arc (180:165:3mm) -- cycle node[left,xshift=-13.5mm,yshift=2mm]{$\beta$}; 
 \draw[gray] (-1.0,0) -- (1.0,0) coordinate (x axis);
 \draw[gray] (0,-1.0) -- (0,1.0) coordinate (y axis);
 \draw (0,0) circle (1cm);
 %\draw[very thick,red] (30:1cm) -- node[left,fill=white] {$\sin \alpha$} (30:1cm |- x axis);
 %\draw[very thick,blue] (30:1cm |- x axis) -- node[below=2pt,fill=white] {$\cos \alpha$} (0,0);
 \draw [dotted] (0,0) -- (60:1cm);
 \draw [dotted] (0,0) -- (300:1cm);
  \draw [dotted,blue] (0,0) -- (35:1cm);
 \draw [dotted,blue] (0,0) -- (275:1cm);
\draw [name path=dotline1][dotted] (60:1cm) -- (300:1cm); 
\draw [name path=line1][dotted] (35:1cm) -- (275:1cm);   
\path [name intersections={of=dotline1 and line1,by=int1}];  
 \filldraw[fill=green!20,draw=green!50!black] (0,0) -- ({1.3mm*cos(60)} ,{1.3mm*sin(60)}) arc (60:35:1.3mm) -- cycle node[left,xshift=6.3mm,yshift=4.8mm]{$\dl$}; 
 %\filldraw[fill=green!20,draw=green!50!black] (int1) -- ++({0 mm} ,-{3 mm} ) arc (270:245:3mm) -- cycle node[left,xshift=-30.5mm,yshift=-5mm]{$\delta$}; 
 \draw [name path=line1b][very thick,blue] (35:1cm) -- (275:1cm);  
  \draw [dotted] (0,0) -- (15:1cm);
 \draw [dotted] (0,0) -- (165:1cm);
\draw [dotted] (0,0) -- (155:1cm); 
\draw [dotted] (0,0) -- (5:1cm); 
  \draw [name path=dotline2][dotted] (15:1cm) -- (165:1cm);
    \draw [name path=line2][dotted] (5:1cm) -- (155:1cm);
   \path [name intersections={of=dotline2 and line2,by=int2}];   
        %\filldraw[fill=red!20,draw=red!50!black] (int2) -- ++(-{3 mm} ,{0 mm} ) arc (180:170:3mm) -- cycle node[left,xshift=-14mm,yshift=-14.25mm]{$\alpha$}; 
          \filldraw[fill=red!20,draw=red!50!black] (0,0) -- (-{3mm*cos(15)} ,{3mm*sin(15)}) arc (165:155:3mm) -- cycle node[left,xshift=-12.5mm,yshift=5.5mm]{$\al$}; 
\draw [name path=line2b][very thick,blue] (5:1cm) -- (155:1cm);
 %\foreach \x/\xtext in {-1, -0.5/-\frac{1}{2}, 1} 
%   \draw (\x cm,1pt) -- (\x cm,-1pt) node[anchor=north,fill=white] {$\xtext$};
 %\foreach \y/\ytext in {-1, -0.5/-\frac{1}{2}, 0.5/\frac{1}{2}, 1} 
  % \draw (1pt,\y cm) -- (-1pt,\y cm) node[anchor=east,fill=white] {$\ytext$};
  \node[label=right:{$g_{R_1}^{-1}(0+\beta-\alpha,z_1)$},shape=circle,fill=black, scale=0.5] at ({(cos(-5))},{(sin(5))}) {};
   \node[label=left:{$g_{R_1}(\pi-\beta-\alpha,z_1)$},shape=circle,fill=red, scale=0.5] at (-{(cos(25))},{(sin(25))}) {};
    \node[label={[label distance=0.5mm]60:$g_{R_2}^{-1}(\frac{\pi}{2}-\rho-\delta,z_2)$},shape=circle,fill=black, scale=0.5] at ({(cos(35))},{(sin(35))}) {};
   \node[label={[label distance=0.5mm]300:$g_{R_2}(\frac{3\pi}{2}+\rho-\delta,z_2)$},shape=circle,fill=red, scale=0.5] at ({(cos(85))},-{(sin(85))}) {};
 \end{tikzpicture}
\caption{Arbitrarily inserted crossed Wilson lines.}
\label{arb}
 \end{figure}

% Instead of spelling out the details of the two aforementioned examples, we shall instead
Let us now study the most general case. 
 We shall show that the four-point function corresponding to any \textit{arbitrary} configuration of crossed Wilson lines has the same expression. Such a configuration, as depicted in Figure \ref{arb}, is determined by four angles, namely $\alpha$, $\beta$, $\gamma$, and $\rho$. 
 The four-point function %to be computed 
 is then
 \begin{equation}\label{big}
\begin{aligned}
&\langle g_{R_1}^{-1}(0+\beta-\alpha,z_1)g_{R_1}(\pi-\beta-\alpha,z_1)\otimes g_{R_2}^{-1}({\pi}/{2}-\rho-\delta,z_2)g_{R_2}({3\pi}/{2}+\rho-\delta,z_2) \rangle\\
=&\mathds{1}+ \frac{\hbar}{z_1-z_2}\frac{1}{2}(\til{\Delta}_{\frac{\pi}{2}+\beta-\rho-\al+\dl}-\til{\Delta}_{-{\frac{\pi}{2}+\beta+\rho-\al+\dl}}+\til{\Delta}_{\frac{\pi}{2}-\beta+\rho-\al+\dl}-\til{\Delta}_{-{\frac{\pi}{2}-\beta-\rho-\al+\dl}})T_{R_1}^a\otimes T_{ R_2a}\\&+O(\hbar^2).
\end{aligned}
\end{equation}
%using \eqref{idn}. 
Here, to ensure single-valuedness of propagators, we require $-\frac{3\pi}{2}<\bt-\rh-\al+\dl<\frac{\pi}{2}$, $-\frac{3\pi}{2}<-\bt+\rh-\al+\dl<\frac{\pi}{2}$, $-\frac{\pi}{2}<\bt+\rh-\al+\dl<\frac{3\pi}{2}$ and $-\frac{\pi}{2}<-\bt-\rh-\al+\dl<\frac{3\pi}{2}$. 
%In addition
However, to ensure that we are considering only crossed Wilson lines, %(to check again) 
we require the stronger conditions $-\frac{\pi}{2}<\bt-\rh-\al+\dl<\frac{\pi}{2}$, $-\frac{\pi}{2}<-\bt+\rh-\al+\dl<\frac{\pi}{2}$, $-\frac{\pi}{2}<\bt+\rh-\al+\dl<\frac{\pi}{2}$ and $-\frac{\pi}{2}<-\bt-\rh-\al+\dl<\frac{\pi}{2}$. 
These conditions in turn allow us to use \eqref{idn}, whereby we find that \eqref{big} is
\begin{equation}
\mathds{1}+ \frac{\hbar}{z_1-z_2}T_{R_1}^a\otimes T_{ R_2a}+O(\hbar^2).
\end{equation}

We have thus shown that topological invariance along $\Sigma$ of the bulk correlation function of two crossed Wilson lines is reflected in the dual four-point function of the boundary 3d ``chiral" WZW model, at least up to order $\hbar$.

\subsection{Crossed Wilson Lines at Order $\hbar^2$}

%Following is written in notes form
Having found the expected %nontrivial leading 
%$\mathcal{O}(\hbar)$ 
order $\hbar$ contribution to the four-point function of arbitrarily inserted crossed Wilson lines in \eqref{big},
%the RHS of \eqref{eqeq},
 we may improve on this and compute higher order contributions by using (free-field) $n$-point functions defined from the generating functional $\eqref{gf}$, for even $n$. We shall demonstrate this explicitly for the order $\hbar^2$ contribution to the correlation function of perpendicular Wilson lines. 
%Although this and higher order contributions are determined in principle by general theorems (as mentioned above), we shall compute it in order to be as explicit as possible.

Firstly, from the generating functional $\eqref{gf}$, the 
(free-field) four-point function 
\begin{equation}\label{per4pt}
\begin{aligned}
&\langle \phi^a({w})\phi^b({x})\phi^c({y})\phi^d({z})\rangle\\=&\frac{\hbar^2}{4}\bigg(\Delta^{ab}({w}-{x})\Delta^{cd}({y}-{z})+\Delta^{ac}({w}-{y})\Delta^{bd}({x}-{z})+\Delta^{ad}({w}-{z})\Delta^{bc}({x}-{y})\bigg)
\end{aligned}
\end{equation}
can be found. 
%(no factors of $2\pi i$ appear here because we use the conventions of Costello-Witten-Yamazaki \cite{CWY} in this subsection).
Expanding the operators in the RHS of \eqref{eqeq} to quadratic order in $\phi$ as
\begin{equation}\label{linordquad}
\begin{aligned}
g^{-1}_{R_1}(0,z_1)g_{R_1}(\pi,z_1)=&\mathds{1}+(\phi_a(\pi,z_1)-\phi_a(0,z_1))T^a_{R_1}+\bigg(-\phi_a(0,z_1)\phi_b(\pi,z_1)
\\&+\frac{1}{2}\phi_a(\pi,z_1)\phi_b(\pi,z_1)+\frac{1}{2}\phi_a(0,z_1)\phi_b(0,z_1)\bigg)T^a_{R_1}T^b_{R_1}+\ldots\\
g^{-1}_{R_2}(\pi/2,z_2)g_{R_2}(3\pi/2,z_2)=&\mathds{1}+(\phi_a(3\pi/2,z_2)-\phi_a(\pi/2,z_2))T^a_{R_2}+\bigg(-\phi_a(\pi/2,z_2)\phi_b(3\pi/2,z_2)
\\&+\frac{1}{2}\phi_a(3\pi/2,z_2)\phi_b(3\pi/2,z_2)+\frac{1}{2}\phi_a(\pi/2,z_2)\phi_b(\pi/2,z_2)\bigg)T^a_{R_2}T^b_{R_2}\\&+\ldots,
\end{aligned}
\end{equation}
we then find via \eqref{per4pt} that, to order $\hbar^2$, \eqref{eqeq} is 
\begin{equation}
\begin{aligned}\label{highero}
&\langle g_{R_1}^{-1}(0,z_1)g_{R_1}(\pi,z_1)\otimes g_{R_2}^{-1}({\pi}/{2},z_2)g_{R_2}({3\pi}/{2},z_2) \rangle\\=&
\mathds{1}+\frac{\hbar}{z_1-z_2}T_{R_1}^a\otimes T_{ R_2a}\\&+
\frac{\hbar^2}{4(z_1-z_2)^2}\Bigg(\TD^{ac}_{0-\frac{\pi}{2}}\TD^{bd}_{\pi-\frac{3\pi}{2}}+\TD^{ad}_{2\pi-\frac{3\pi}{2}}\TD^{bc}_{\pi-\frac{\pi}{2}}-\frac{1}{2}\bigg(\TD^{ac}_{2\pi-\frac{3\pi}{2}}\TD^{bd}_{\pi-\frac{3\pi}{2}}+\TD^{ad}_{2\pi-\frac{3\pi}{2}}\TD^{bc}_{\pi-\frac{3\pi}{2}}\bigg)\\&~~~~~~~~~~~~~~~~~~~~-\frac{1}{2}\bigg(\TD^{ac}_{0-\frac{\pi}{2}}\TD^{bd}_{\pi-\frac{\pi}{2}}+\TD^{ad}_{0-\frac{\pi}{2}}\TD^{bc}_{\pi-\frac{\pi}{2}}\bigg)-\frac{1}{2}\bigg(\TD^{ac}_{\pi-\frac{\pi}{2}}\TD^{bd}_{\pi-\frac{3\pi}{2}}+\TD^{ad}_{\pi-\frac{3\pi}{2}}\TD^{bc}_{\pi-\frac{\pi}{2}}\bigg)\\&~~~~~~~~~~~~~~~~~~~~-\frac{1}{2}\bigg(\TD^{ac}_{0-\frac{\pi}{2}}\TD^{bd}_{2\pi-\frac{3\pi}{2}}+\TD^{ad}_{2\pi-\frac{3\pi}{2}}\TD^{bc}_{0-\frac{\pi}{2}}\bigg)+\frac{1}{4}\bigg(\TD^{ac}_{\pi-\frac{3\pi}{2}}\TD^{bd}_{\pi-\frac{3\pi}{2}}+\TD^{ad}_{\pi-\frac{3\pi}{2}}\TD^{bc}_{\pi-\frac{3\pi}{2}}\bigg)\\&~~~~~~~~~~~~~~~~~~~~+\frac{1}{4}\bigg(\TD^{ac}_{\pi-\frac{\pi}{2}}\TD^{bd}_{\pi-\frac{\pi}{2}}+\TD^{ad}_{\pi-\frac{\pi}{2}}\TD^{bc}_{\pi-\frac{\pi}{2}}\bigg)+\frac{1}{4}\bigg(\TD^{ac}_{2\pi-\frac{3\pi}{2}}\TD^{bd}_{2\pi-\frac{3\pi}{2}}+\TD^{ad}_{2\pi-\frac{3\pi}{2}}\TD^{bc}_{2\pi-\frac{3\pi}{2}}\bigg)\\&~~~~~~~~~~~~~~~~~~~~+\frac{1}{4}\bigg(\TD^{ac}_{0-\frac{\pi}{2}}\TD^{bd}_{0-\frac{\pi}{2}}+\TD^{ad}_{0-\frac{\pi}{2}}\TD^{bc}_{0-\frac{\pi}{2}}\bigg)\Bigg)T^a_{R_1}T^b_{R_1}\otimes T^c_{R_2}T^d_{R_2}+\mathcal{O}(\hbar^3)
\\=&
\mathds{1}+\frac{\hbar}{z_1-z_2}T_{R_1}^a\otimes T_{ R_2a}+\frac{\hbar^2}{4(z_1-z_2)^2}\big(T^a_{R_1}T^b_{R_1}\otimes T_{R_2a}T_{R_2b} + T^a_{R_1}T^b_{R_1}\otimes T_{R_2b}T_{R_2a}\big)+\mathcal{O}(\hbar^3),
\end{aligned}
\end{equation}
where we have used the notation $\TD^{ab}_{\vphi}=\TD_{\vphi}\dl^{ab}$ for brevity.

In a similar manner, one can compute contributions to the $R$-matrix of order $\hbar^3$ and above. Note that %these 
contributions at order $\hbar^2$ and above are not expected to remain invariant under moves of the local boundary operators that correspond to rotations and translations of the bulk Wilson lines, due to the framing anomaly that arises in the bulk theory at order $\hbar^2$ for non-perpendicular Wilson lines \cite{CWY}. This framing anomaly ought to be computable in our boundary WZW model as well, by taking into account its interaction terms %(more accurate to say interaction terms, because such terms come from WZW kinetic term) 
when computing correlation functions of local operators. 

 A slightly more involved calculation shows that the result of \eqref{highero} holds, modulo the framing anomaly, for arbitrarily inserted Wilson lines (as depicted in Figure \ref{arb}), assuming the same constraints on the angles given below \eqref{big}.
\subsection{Parallel Wilson Lines}
%We shall now consider uncrossed Wilson lines and the 
%The OPEs of uncrossed Wilson lines provide one method of deriving the Yangian algebra in 4d Chern-Simons theory. In this subsection, we shall consider such correlation functions of  uncrossed Wilson lines and show how they are captured by correlation functions of boundary operators. We shall focus on the free-field limit, at order $\hbar$, and retrieve the expected behaviour in this regime.
The OPEs of parallel Wilson lines in 4d Chern-Simons theory do not have the same
singular behaviour as correlation functions of crossed Wilson lines. In this subsection, we
shall consider such correlation functions of parallel Wilson lines and show how they are
captured by correlation functions of boundary operators. We shall focus on the free-field limit, at order $\hbar$, and retrieve the expected behaviour in this regime.

\begin{figure}[!h]
\centering
\begin{tikzpicture}[scale=3]
% \draw[step=.5cm, gray, very thin] (-1.2,-1.2) grid (1.2,1.2); 
 %\filldraw[fill=green!20,draw=green!50!black] (0,0) -- ({(cos(60))*3 mm} ,{(sin(60))*3 mm} ) arc (60:90:3mm) -- cycle node[above,xshift=3mm,yshift=9mm]{$\delta$}; 
 \draw[gray] (-1.0,0) -- (1.0,0) coordinate (x axis);
 \draw[gray] (0,-1.0) -- (0,1.0) coordinate (y axis);
 \draw (0,0) circle (1cm);
 %\draw[very thick,red] (30:1cm) -- node[left,fill=white] {$\sin \alpha$} (30:1cm |- x axis);
 %\draw[very thick,blue] (30:1cm |- x axis) -- node[below=2pt,fill=white] {$\cos \alpha$} (0,0);
% \draw [very thick,blue] (0,0) -- (90:1cm);
 %\draw [very thick,blue] (0,0) -- (270:1cm);
  %\draw [very thick,blue] (0,0) -- (0:1cm);
 %\draw [very thick,blue] (0,0) -- (180:1cm);
 \draw [name path=line1b][very thick,blue] (270:1cm) -- (0:1cm); 
  \draw [name path=line1b][very thick,blue] (180:1cm) -- (90:1cm); 
 %\foreach \x/\xtext in {-1, -0.5/-\frac{1}{2}, 1} 
%   \draw (\x cm,1pt) -- (\x cm,-1pt) node[anchor=north,fill=white] {$\xtext$};
 %\foreach \y/\ytext in {-1, -0.5/-\frac{1}{2}, 0.5/\frac{1}{2}, 1} 
  % \draw (1pt,\y cm) -- (-1pt,\y cm) node[anchor=east,fill=white] {$\ytext$};
  \node[label=right:{$g^{-1}_{R_1}(0,z_1,\zb_1)$},shape=circle,fill=black, scale=0.5] at (1,0) {};
   \node[label=left:{$g_{R_2}(\pi,z_2,\zb_2)$},shape=circle,fill=red, scale=0.5] at (-1,0) {};
    \node[label=above:{$g^{-1}_{R_2}(\frac{\pi}{2},z_2,\zb_2)$},shape=circle,fill=black, scale=0.5] at ({(cos(90))},{(sin(90))}) {};
   \node[label=below:{$g_{R_1}(\frac{3\pi}{2},z_1,\zb_1)$},shape=circle,fill=red, scale=0.5] at (-{(cos(90))},-{(sin(90))}) {};
 \end{tikzpicture}
 \caption{Parallel Wilson lines on $D$.}
 \label{uncr}
\end{figure}
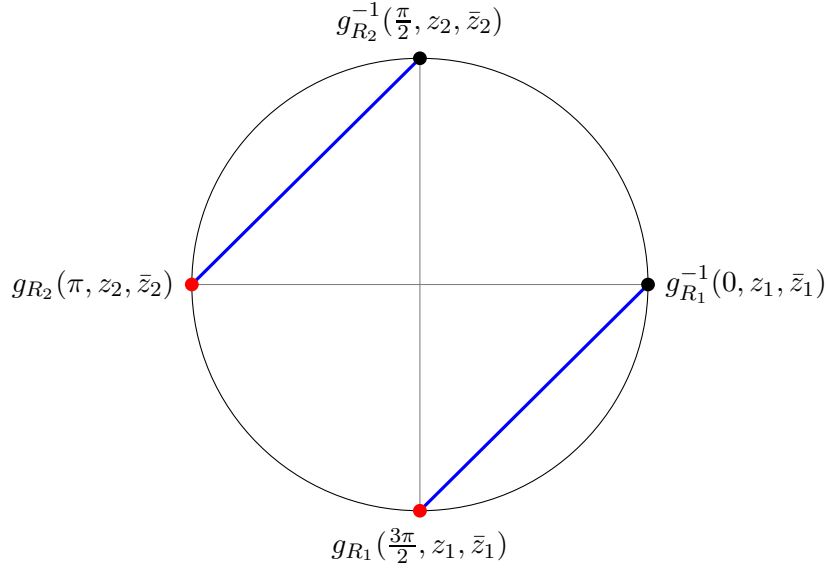

 Using \eqref{hol}, the correlation function of the operators we are interested in (depicted in Figure \ref{uncr}) is
 \begin{equation}\label{eqeq2}
 \begin{aligned}
 &\langle \mathcal{P} e^{\int_{\varphi=3\pi/2}^{\varphi=0} \cA_{R_1}(z_1,\zb_1)}\otimes \mathcal{P} e^{\int^{\varphi=\pi/2}_{\varphi=\pi}\cA_{R_2}(z_2,\zb_2)}\rangle \\=& \langle g_{R_1}^{-1}(0,z_1,\zb_1)g_{R_1}(3\pi/2,z_1,\zb_1)\otimes g_{R_2}^{-1}({\pi}/{2},z_2,\zb_2)g_{R_2}({\pi},z_2,\zb_2) \rangle,
 \end{aligned}
   \end{equation}
(note the difference from \eqref{eqeq} in ordering of the boundary operators when $z_1=z_2$ and $R_1=R_2$). 

Expanding each operator to linear order in $\phi$ and keeping only terms of quadratic or lower order in the fields (as in the discussion below \eqref{linord}), we have 
\begin{equation}\label{res1uncr}
\begin{aligned}
&\mathds{1}+\langle \phi_a(2\pi,z_1)\phi_c(\pi/2,z_2)\rangle T_{R_1}^a\otimes T_{R_2}^c-\langle \phi_a(2\pi,z_1)\phi_c(\pi,z_2) \rangle T_{R_1}^a\otimes T_{R_2}^c\\-&\langle \phi_a(3\pi/2,z_1)\phi_c(\pi/2,z_2) \rangle T_{R_1}^a\otimes T_{R_2}^c+\langle \phi_a(3\pi/2,z_1)\phi_c(\pi,z_2)\rangle T_{R_1}^a\otimes T_{R_2}^c+O(\hbar^2)\\
=&\mathds{1}-\frac{\hbar}{2}\delta_{ac} \frac{1}{z_1-z_2}\til{\Delta}_{\frac{3\pi}{2}}T_{R_1}^a\otimes T_{R_2}^c+\frac{\hbar}{2}\delta_{ac} \frac{1}{z_1-z_2}\til{\Delta}_{{\pi}}T_{R_1}^a\otimes T_{R_2}^c\\&+\frac{\hbar}{2}\delta_{ac} \frac{1}{z_1-z_2}\til{\Delta}_{{\pi}}T_{R_1}^a\otimes T_{R_2}^c-\frac{\hbar}{2}\delta_{ac} \frac{1}{z_1-z_2}\til{\Delta}_{\frac{\pi}{2}}T_{R_1}^a\otimes T_{R_2}^c+O(\hbar^2)\\
%=&\mathds{1}+\frac{1}{2\pi i}\frac{\hbar}{z_1-z_2}(\til{\Delta}_{\frac{\pi}{2}}-%\til{\Delta}_{{\frac{\pi}{2}}})T_{R_1}^a\otimes T_{ R_2a}+O(\hbar^2)\\
=&\mathds{1} +O(\hbar^2),
\end{aligned}
\end{equation}
which is non-singular for $z_1=z_2$ at order $\hbar$, as expected. 
%(Mention independence of ordering of operators in the bulk.)
Here, we have used the previously derived fact that $\til{\Delta}_{\frac{\pi}{2}}=\frac{1}{2}$, as well as $\til{\Delta}_{\frac{3\pi}{2}}=\frac{1}{2}$, which follows from 
\eqref{ff}, \eqref{gd} and \eqref{gd3}, and 
\begin{equation}\label{multpiuncr}
\begin{aligned}
\widetilde{\Delta}_{\pi}&=\frac{1}{2\pi}\bigg(\sum_{k=1}^{\infty}\frac{(-1)^k}{ik}+\pi+\sum_{k=-\infty}^{-1}\frac{(-1)^k}{ik}\bigg)\\
&=\frac{1}{2\pi}\bigg(i\textrm{ln } 2 +\pi - i\textrm{ln } 2  \bigg)\\
&=\frac{1}{2}.
\end{aligned}
\end{equation}
Also, note that we must define the propagator here with a different branch cut from that of previous sections, namely, from $r=0$ to ($r=R$, $\vphi=0$). This effectively restricts $\vphi$ in \eqref{mult} to take values in $(0,{2\pi})$.
\subsection{Three Wilson Lines}

\begin{figure}[!h]
\centering
\begin{tikzpicture}[scale=3]
% \draw[step=.5cm, gray, very thin] (-1.2,-1.2) grid (1.2,1.2); 
 %\filldraw[fill=blue!20,draw=blue!50!black] (0,0) -- ({(cos(60))*1.3 mm} ,{(sin(60))*1.3 mm} ) arc (60:90:1.3mm) -- cycle node[above,xshift=1.5mm,yshift=2.9mm]{$\rho$}; 
  %\filldraw[fill=yellow!20,draw=yellow!50!black] (0,0) -- ({-3 mm} ,{0 mm} ) arc (180:165:3mm) -- cycle node[left,xshift=-13.5mm,yshift=2mm]{$\beta$}; 
 \draw[gray] (-1.0,0) -- (1.0,0) coordinate (x axis);
 \draw[gray] (0,-1.0) -- (0,1.0) coordinate (y axis);
 \draw (0,0) circle (1cm);
 %\draw[very thick,red] (30:1cm) -- node[left,fill=white] {$\sin \alpha$} (30:1cm |- x axis);
 %\draw[very thick,blue] (30:1cm |- x axis) -- node[below=2pt,fill=white] {$\cos \alpha$} (0,0);
% \draw [dotted] (0,0) -- (60:1cm);
% \draw [dotted] (0,0) -- (300:1cm);
 % \draw [dotted,blue] (0,0) -- (35:1cm);
 %\draw [dotted,blue] (0,0) -- (275:1cm);
%\draw [name path=dotline1][dotted] (60:1cm) -- (300:1cm); 
%\draw [name path=line1][dotted] (35:1cm) -- (275:1cm);   
%\path [name intersections={of=dotline1 and line1,by=int1}];  
% \filldraw[fill=green!20,draw=green!50!black] (0,0) -- ({1.3mm*cos(60)} ,{1.3mm*sin(60)}) arc (60:35:1.3mm) -- cycle node[left,xshift=6.3mm,yshift=4.8mm]{$\dl$}; 
 %\filldraw[fill=green!20,draw=green!50!black] (int1) -- ++({0 mm} ,-{3 mm} ) arc (270:245:3mm) -- cycle node[left,xshift=-30.5mm,yshift=-5mm]{$\delta$}; 
 \draw [name path=line1b][very thick,blue] (15:1cm) -- (193:1cm);  
  %\draw [dotted] (0,0) -- (15:1cm);
 %\draw [dotted] (0,0) -- (165:1cm);
%\draw [dotted] (0,0) -- (155:1cm); 
%\draw [dotted] (0,0) -- (5:1cm); 
  %\draw [name path=dotline2][dotted] (15:1cm) -- (165:1cm);
   % \draw [name path=line2][dotted] (5:1cm) -- (155:1cm);
   %\path [name intersections={of=dotline2 and line2,by=int2}];   
        %\filldraw[fill=red!20,draw=red!50!black] (int2) -- ++(-{3 mm} ,{0 mm} ) arc (180:170:3mm) -- cycle node[left,xshift=-14mm,yshift=-14.25mm]{$\alpha$}; 
         % \filldraw[fill=red!20,draw=red!50!black] (0,0) -- (-{3mm*cos(15)} ,{3mm*sin(15)}) arc (165:155:3mm) -- cycle node[left,xshift=-12.5mm,yshift=5.5mm]{$\al$}; 
\draw [name path=line2b][very thick,blue] (-15:1cm) -- (145:1cm);
\draw [name path=line2b][very thick,blue] (92:1cm) -- (260:1cm);
 %\foreach \x/\xtext in {-1, -0.5/-\frac{1}{2}, 1} 
%   \draw (\x cm,1pt) -- (\x cm,-1pt) node[anchor=north,fill=white] {$\xtext$};
 %\foreach \y/\ytext in {-1, -0.5/-\frac{1}{2}, 0.5/\frac{1}{2}, 1} 
  % \draw (1pt,\y cm) -- (-1pt,\y cm) node[anchor=east,fill=white] {$\ytext$};
  \node[label=right:{$g_{R_1}^{-1}(0+\beta-\rho,z_1)$},shape=circle,fill=black, scale=0.5] at ({(cos(-15))},{(sin(15))}) {};
     \node[label=right:{$\g_{R_3}(0+\gamma-\zeta,z_3)/\atop g_{R_3}(\frac{3\pi}{2}+\gamma'-\zeta',z_3)$},shape=circle,fill=red, scale=0.5] at ({(cos(15))},{-(sin(15))}) {};
   \node[label=left:{$g_{R_1}(\pi-\beta-\rho,z_1)$},shape=circle,fill=red, scale=0.5] at ({(cos(193))},{(sin(193))}) {};
      \node[label=left:{$g_{R_3}^{-1}(\pi-\gamma-\zeta,z_3)/\atop g_{R_3}^{-1}(\frac{\pi}{2}-\gamma'-\zeta',z_3)$},shape=circle,fill=black, scale=0.5] at ({(cos(145))},{(sin(145))}) {};
    \node[label={[label distance=0.5mm]60:$g_{R_2}^{-1}(\frac{\pi}{2}-\delta-\al,z_2)$},shape=circle,fill=black, scale=0.5] at ({(cos(92))},{(sin(92))}) {};
   \node[label={[label distance=0.5mm]300:$g_{R_2}(\frac{3\pi}{2}+\dl-\al,z_2)$},shape=circle,fill=red, scale=0.5] at ({(cos(260))},{(sin(260))}) {};
 \end{tikzpicture}
\caption{Three Wilson lines.}
\label{arb3}
 \end{figure}
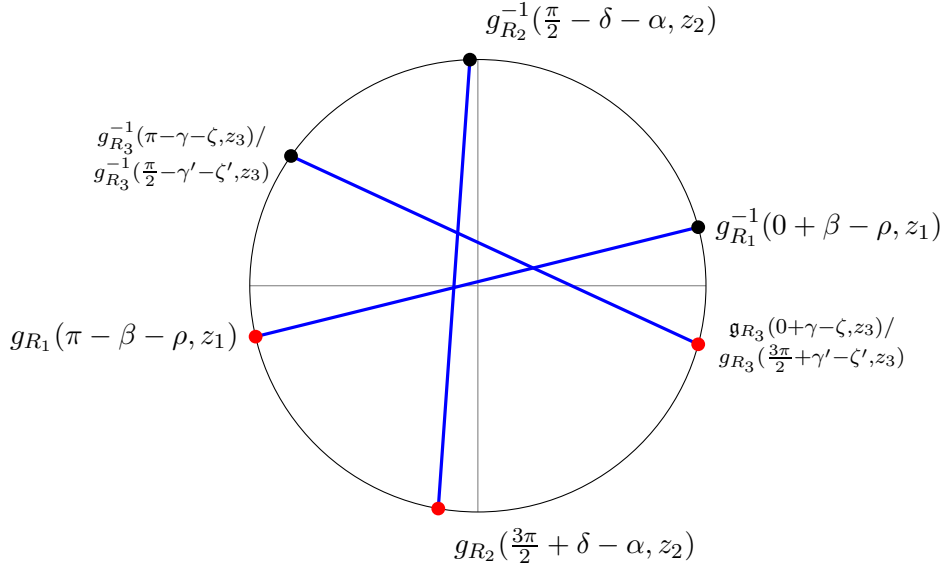

We next consider correlation functions of three Wilson lines, all crossing each other %in the configuration given 
but otherwise inserted arbitrarily (c.f. Figure \ref{arb3}), which corresponds to the following boundary correlator:
\begin{equation}\label{eqeq3}
 \begin{aligned}
 & \langle g_{R_1}^{-1}(0+\bt-\rho)g_{R_1}(\pi-\bt-\rho)\otimes g_{R_2}^{-1}(\frac{\pi}{2}-\dl-\al)g_{R_2}(\frac{3\pi}{2}+\dl-\al) \\  & \quad \otimes g_{R_3}^{-1}(\pi-\gm-\zeta)g_{R_3}(0+\gm-\zeta)\rangle,
 \end{aligned}
   \end{equation}
where the dependence on $\C$ has been suppressed for brevity.
Expanding each operator in \eqref{eqeq3} to linear order in $\phi$ we find
\begin{equation}
\begin{aligned}
&\mathds{1}+\big(\langle \phi_a(0+\bt+\rho,z_1)\phi_c(\pi/2-\delta-\al,z_2)\rangle -\langle \phi_a(\pi-\bt-\rho,z_1)\phi_c(\pi/2-\delta-\al,z_2)\rangle \\&-\langle \phi_a(2\pi+\bt-\rho,z_1)\phi_c(3\pi/2+\delta-\al,z_2)\rangle +\langle \phi_a(\pi-\bt-\rho,z_1)\phi_c(3\pi/2+\delta-\al,z_2)\rangle \big) T_{R_1}^a\otimes T_{R_2}^c \otimes \mathds{1} \\
&+\big(\langle \phi_a(0+\bt-\rho,z_1)\phi_c(\pi/2-\gm'-\zt',z_3)\rangle -\langle \phi_a(2\pi+\bt-\rho,z_1)\phi_c(3\pi/2+\gm'-\zt',z_3)\rangle \\&-\langle \phi_a(\pi-\bt-\rho,z_1)\phi_c(\pi/2-\gm'-\zt',z_3)\rangle +\langle \phi_a(\pi-\bt-\rho,z_1)\phi_c(3\pi/2+\gm'-\zt',z_3)\rangle \big) T_{R_1}^a\otimes \mathds{1} \otimes T_{R_3}^c  \\
&+\big(\langle \phi_a(\pi/2-\dl-\al,z_2)\phi_c(\pi-\gm-\zt,z_3)\rangle -\langle \phi_a(\pi/2-\dl-\al,z_2)\phi_c(0+\gm-\zt,z_3)\rangle \\&-\langle \phi_a(3\pi/2+\dl-\al,z_2)\phi_c(\pi-\gm-\zt,z_3)\rangle +\langle \phi_a(3\pi/2+\dl-\al,z_2)\phi_c(2\pi+\gm-\zt,z_3)\rangle \big) \mathds{1} \otimes T_{R_2}^a\otimes T_{R_3}^c \\&+O(\hbar^2)\\
&=\mathds{1}+\frac{\hbar}{z_1-z_2}\frac{1}{2}\bigg(\frac{1}{2}+\frac{1}{2}+\frac{1}{2}+\frac{1}{2}\bigg)T_{R_1}^a\otimes T_{R_2a} \otimes \mathds{1}+\frac{\hbar}{z_1-z_3}\frac{1}{2}\bigg(\frac{1}{2}+\frac{1}{2}+\frac{1}{2}+\frac{1}{2}\bigg)T_{R_1}^a\otimes \mathds{1}\otimes T_{R_3a} \\&-\frac{\hbar}{z_3-z_2}\frac{1}{2}\bigg(\frac{1}{2}+\frac{1}{2}+\frac{1}{2}+\frac{1}{2}\bigg)\mathds{1} \otimes T_{R_2}^a\otimes T_{R_3a}+O(\hbar^2)\\ &=\mathds{1}+\frac{\hbar}{z_1-z_2}T_{R_1}^a\otimes T_{R_2a} \otimes \mathds{1}+\frac{\hbar}{z_1-z_3} T_{R_1}^a\otimes \mathds{1}\otimes T_{R_3a} +\frac{\hbar}{z_2-z_3}\mathds{1} \otimes T_{R_2}^a\otimes T_{R_3a}+O(\hbar^2),
\end{aligned}
\end{equation}
where various constraints on the angles are necessary for single-valuedness of propagators 	and to ensure that the Wilson lines are all crossed. Once again, there is agreement with the bulk 4d Chern-Simons computation.

\begin{figure}[!h]
	\centering	
			\captionsetup[subfigure]{oneside,margin={0.1cm,0cm}}
			\subfloat[]{
			\begin{tikzpicture}[scale=2.3]\label{9a}
			% \draw[step=.5cm, gray, very thin] (-1.2,-1.2) grid (1.2,1.2); 
			%\filldraw[fill=blue!20,draw=blue!50!black] (0,0) -- ({(cos(60))*1.3 mm} ,{(sin(60))*1.3 mm} ) arc (60:90:1.3mm) -- cycle node[above,xshift=1.5mm,yshift=2.9mm]{$\rho$}; 
			%\filldraw[fill=yellow!20,draw=yellow!50!black] (0,0) -- ({-3 mm} ,{0 mm} ) arc (180:165:3mm) -- cycle node[left,xshift=-13.5mm,yshift=2mm]{$\beta$}; 
			\draw[gray] (-1,0) -- (1,0) coordinate (x axis);
			\draw[gray] (0,-1) -- (0,1) coordinate (y axis);
			\draw (0,0) circle (1cm);
			%\draw[very thick,red] (30:1cm) -- node[left,fill=white] {$\sin \alpha$} (30:1cm |- x axis);
			%\draw[very thick,blue] (30:1cm |- x axis) -- node[below=2pt,fill=white] {$\cos \alpha$} (0,0);
			% \draw [dotted] (0,0) -- (60:1cm);
			% \draw [dotted] (0,0) -- (300:1cm);
			% \draw [dotted,blue] (0,0) -- (35:1cm);
			%\draw [dotted,blue] (0,0) -- (275:1cm);
			%\draw [name path=dotline1][dotted] (60:1cm) -- (300:1cm); 
			%\draw [name path=line1][dotted] (35:1cm) -- (275:1cm);   
			%\path [name intersections={of=dotline1 and line1,by=int1}];  
			% \filldraw[fill=green!20,draw=green!50!black] (0,0) -- ({1.3mm*cos(60)} ,{1.3mm*sin(60)}) arc (60:35:1.3mm) -- cycle node[left,xshift=6.3mm,yshift=4.8mm]{$\dl$}; 
			%\filldraw[fill=green!20,draw=green!50!black] (int1) -- ++({0 mm} ,-{3 mm} ) arc (270:245:3mm) -- cycle node[left,xshift=-30.5mm,yshift=-5mm]{$\delta$}; 
			\draw [name path=line1b][very thick,blue] (30:1 cm) -- (210:1 cm);  
			%\draw [dotted] (0,0) -- (15:1cm);
			%\draw [dotted] (0,0) -- (165:1cm);
			%\draw [dotted] (0,0) -- (155:1cm); 
			%\draw [dotted] (0,0) -- (5:1cm); 
			%\draw [name path=dotline2][dotted] (15:1cm) -- (165:1cm);
			% \draw [name path=line2][dotted] (5:1cm) -- (155:1cm);
			%\path [name intersections={of=dotline2 and line2,by=int2}];   
			%\filldraw[fill=red!20,draw=red!50!black] (int2) -- ++(-{3 mm} ,{0 mm} ) arc (180:170:3mm) -- cycle node[left,xshift=-14mm,yshift=-14.25mm]{$\alpha$}; 
			% \filldraw[fill=red!20,draw=red!50!black] (0,0) -- (-{3mm*cos(15)} ,{3mm*sin(15)}) arc (165:155:3mm) -- cycle node[left,xshift=-12.5mm,yshift=5.5mm]{$\al$}; 
			\draw [name path=line2b][very thick,blue] (-35:1 cm) -- (145:1 cm);
			\draw [name path=line2b][very thick,blue] (105:1 cm) -- (250:1 cm);
			%\foreach \x/\xtext in {-1, -0.5/-\frac{1}{2}, 1} 
			%   \draw (\x cm,1pt) -- (\x cm,-1pt) node[anchor=north,fill=white] {$\xtext$};
			%\foreach \y/\ytext in {-1, -0.5/-\frac{1}{2}, 0.5/\frac{1}{2}, 1} 
			% \draw (1pt,\y cm) -- (-1pt,\y cm) node[anchor=east,fill=white] {$\ytext$};
			\node[shape=circle,fill=black, scale=0.5 ] at ({(cos(-30))},{(sin(30))}) {};
			\node[shape=circle,fill=red, scale=0.5 ] at ({(cos(35))},{-(sin(35))}) {};
			\node[shape=circle,fill=red, scale=0.5] at ({(cos(210))},{(sin(210))}) {};
			\node[shape=circle,fill=black, scale=0.5] at ({(cos(145))},{(sin(145))}) {};
			\node[shape=circle,fill=black, scale=0.5] at ({(cos(105))},{(sin(105))}) {};
			\node[shape=circle,fill=red, scale=0.5] at ({(cos(250))},{(sin(250))}) {};
			\end{tikzpicture}
			}			
			\subfloat[]{
			\begin{tikzpicture}[scale=2.3]\label{9c}
			% \draw[step=.5cm, gray, very thin] (-1.2,-1.2) grid (1.2,1.2); 
			%\filldraw[fill=blue!20,draw=blue!50!black] (0,0) -- ({(cos(60))*1.3 mm} ,{(sin(60))*1.3 mm} ) arc (60:90:1.3mm) -- cycle node[above,xshift=1.5mm,yshift=2.9mm]{$\rho$}; 
			%\filldraw[fill=yellow!20,draw=yellow!50!black] (0,0) -- ({-3 mm} ,{0 mm} ) arc (180:165:3mm) -- cycle node[left,xshift=-13.5mm,yshift=2mm]{$\beta$}; 
			\draw[gray] (-1,0) -- (1,0) coordinate (x axis);
			\draw[gray] (0,-1) -- (0,1) coordinate (y axis);
			\draw (0,0) circle (1cm);
			%\draw[very thick,red] (30:1cm) -- node[left,fill=white] {$\sin \alpha$} (30:1cm |- x axis);
			%\draw[very thick,blue] (30:1cm |- x axis) -- node[below=2pt,fill=white] {$\cos \alpha$} (0,0);
			% \draw [dotted] (0,0) -- (60:1cm);
			% \draw [dotted] (0,0) -- (300:1cm);
			% \draw [dotted,blue] (0,0) -- (35:1cm);
			%\draw [dotted,blue] (0,0) -- (275:1cm);
			%\draw [name path=dotline1][dotted] (60:1cm) -- (300:1cm); 
			%\draw [name path=line1][dotted] (35:1cm) -- (275:1cm);   
			%\path [name intersections={of=dotline1 and line1,by=int1}];  
			% \filldraw[fill=green!20,draw=green!50!black] (0,0) -- ({1.3mm*cos(60)} ,{1.3mm*sin(60)}) arc (60:35:1.3mm) -- cycle node[left,xshift=6.3mm,yshift=4.8mm]{$\dl$}; 
			%\filldraw[fill=green!20,draw=green!50!black] (int1) -- ++({0 mm} ,-{3 mm} ) arc (270:245:3mm) -- cycle node[left,xshift=-30.5mm,yshift=-5mm]{$\delta$}; 
			\draw [name path=line1b][very thick,blue] (30:1 cm) -- (210:1 cm);  
			%\draw [dotted] (0,0) -- (15:1cm);
			%\draw [dotted] (0,0) -- (165:1cm);
			%\draw [dotted] (0,0) -- (155:1cm); 
			%\draw [dotted] (0,0) -- (5:1cm); 
			%\draw [name path=dotline2][dotted] (15:1cm) -- (165:1cm);
			% \draw [name path=line2][dotted] (5:1cm) -- (155:1cm);
			%\path [name intersections={of=dotline2 and line2,by=int2}];   
			%\filldraw[fill=red!20,draw=red!50!black] (int2) -- ++(-{3 mm} ,{0 mm} ) arc (180:170:3mm) -- cycle node[left,xshift=-14mm,yshift=-14.25mm]{$\alpha$}; 
			% \filldraw[fill=red!20,draw=red!50!black] (0,0) -- (-{3mm*cos(15)} ,{3mm*sin(15)}) arc (165:155:3mm) -- cycle node[left,xshift=-12.5mm,yshift=5.5mm]{$\al$}; 
			\draw [name path=line2b][very thick,blue] (-35:1 cm) -- (145:1 cm);
			\draw [name path=line2b][very thick,blue] (70:1 cm) -- (285:1 cm);
			%\foreach \x/\xtext in {-1, -0.5/-\frac{1}{2}, 1} 
			%   \draw (\x cm,1pt) -- (\x cm,-1pt) node[anchor=north,fill=white] {$\xtext$};
			%\foreach \y/\ytext in {-1, -0.5/-\frac{1}{2}, 0.5/\frac{1}{2}, 1} 
			% \draw (1pt,\y cm) -- (-1pt,\y cm) node[anchor=east,fill=white] {$\ytext$};
			\node[shape=circle,fill=black, scale=0.5 ] at ({(cos(-30))},{(sin(30))}) {};
			\node[shape=circle,fill=red, scale=0.5 ] at ({(cos(35))},{-(sin(35))}) {};
			\node[shape=circle,fill=red, scale=0.5] at ({(cos(210))},{(sin(210))}) {};
			\node[shape=circle,fill=black, scale=0.5] at ({(cos(145))},{(sin(145))}) {};
			\node[shape=circle,fill=black, scale=0.5] at ({(cos(70))},{(sin(70))}) {};
			\node[shape=circle,fill=red, scale=0.5] at ({(cos(285))},{(sin(285))}) {}; \end{tikzpicture}
			}
	\caption{The Yang-Baxter equation is realized by moving a Wilson line across the intersection of two other Wilson lines.}
	\label{threed}
\end{figure}
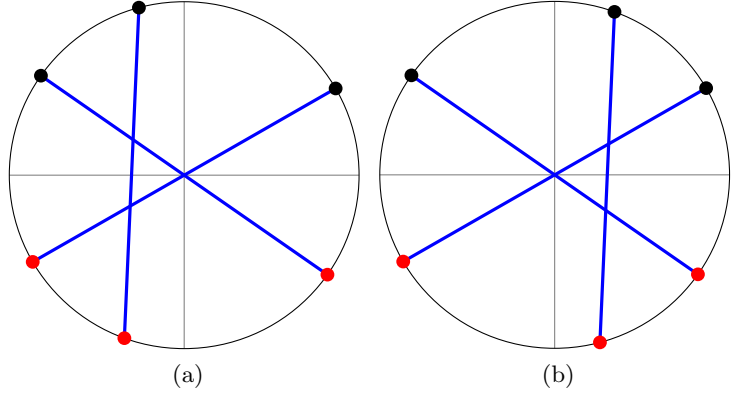

 We may further compute the correlation function \eqref{eqeq3} to order $\hbar^2$ by expanding each operator in \eqref{eqeq3} to quadratic order in $\phi$. Doing so, we find the following expression (modulo the framing anomaly):
\begin{equation}\label{threee}
\begin{aligned}
&\mathds{1}+\frac{\hbar}{z_1-z_2}T_{R_1}^a\otimes T_{R_2a} \otimes \mathds{1}+\frac{\hbar}{z_1-z_3} T_{R_1}^a\otimes \mathds{1}\otimes T_{R_3a} +\frac{\hbar}{z_2-z_3}\mathds{1} \otimes T_{R_2}^a\otimes T_{R_3a}\\&+\frac{\hbar^2}{4(z_1-z_2)^2}\big(T^a_{R_1}T^b_{R_1}\otimes T_{R_2a}T_{R_2b}\otimes \mathds{1} + T^a_{R_1}T^b_{R_1}\otimes T_{R_2b}T_{R_2a}\otimes \mathds{1})
\\ &+\frac{\hbar^2}{4(z_1-z_3)^2}\big(T^a_{R_1}T^b_{R_1}\otimes \mathds{1}\otimes T_{R_3a}T_{R_3b} + T^a_{R_1}T^b_{R_1}\otimes \mathds{1}\otimes T_{R_3b}T_{R_3a} \big)\\ &+\frac{\hbar^2}{4(z_2-z_3)^2}\big(\mathds{1}\otimes T^a_{R_2}T^b_{R_2}\otimes T_{R_3a}T_{R_3b} +  \mathds{1}\otimes T^a_{R_1}T^b_{R_1}\otimes T_{R_2b}T_{R_2a})\\&+\frac{\hbar^2}{2(z_1-z_2)(z_1-z_3)}\big(T^a_{R_1}T^b_{R_1}\otimes T_{R_2a}\otimes T_{R_3b} + T^a_{R_1}T^b_{R_1}\otimes T_{R_2b}\otimes T_{R_3a})\\&+\frac{\hbar^2}{2(z_1-z_2)(z_2-z_3)}\big( T^a_{R_1}\otimes T_{R_2a}T_{R_2b}\otimes T^b_{R_3} + T^a_{R_1}\otimes T^b_{R_2}T_{R_2a}\otimes T_{R_3b} )\\&+\frac{\hbar^2}{2(z_1-z_3)(z_2-z_3)}\big(T^a_{R_1}\otimes T^b_{R_2}\otimes T_{R_3a}T_{R_3b} + T^a_{R_1}\otimes T^b_{R_2}\otimes T_{R_3b}T_{R_3a}) +O(\hbar^3).
\end{aligned}
\end{equation}
This result agrees with the bulk 4d Chern-Simons computation. 
%is of expected form, except in the last three lines. This can be remedied by using 
To see this, let us consider the configurations in Figure \ref{threed}. 
%For these two configurations, all the angles that determine the positions of the Wilson lines (as depicted in Figure \ref{arb3}) except $\delta$ are the same. 
From the bulk theory, we know that the equivalence of these two configurations gives rise to the Yang-Baxter equation
\begin{equation}\label{YB}
\til{R}_{12}\til{R}_{13}\til{R}_{23}=\til{R}_{23}\til{R}_{13}\til{R}_{12}, 
\end{equation}
%use the identity 
where
\begin{equation}
\begin{aligned}
\til{R}_{12}=&\mathds{1}
+\frac{\hbar}{z_1-z_2}T_{R_1}^a\otimes T_{ R_2a}\otimes \mathds{1}\\&+\frac{\hbar^2}{4(z_1-z_2)^2}\big(T^a_{R_1}T^b_{R_1}\otimes T_{R_2a}T_{R_2b}\otimes \mathds{1} + T^a_{R_1}T^b_{R_1}\otimes T_{R_2b}T_{R_2a}\otimes \mathds{1}\big)+\mathcal{O}(\hbar^3),\\
\til{R}_{13}=&\mathds{1}
+\frac{\hbar}{z_1-z_3}T_{R_1}^a\otimes \mathds{1}\otimes T_{ R_3a}\\&+\frac{\hbar^2}{4(z_1-z_3)^2}\big(T^a_{R_1}T^b_{R_1}\otimes \mathds{1}\otimes T_{R_3a}T_{R_3b} + T^a_{R_1}T^b_{R_1}\otimes \mathds{1}\otimes T_{R_3b}T_{R_3a}\big)+\mathcal{O}(\hbar^3),\\\til{R}_{23}=&\mathds{1}
+\frac{\hbar}{z_2-z_3}\mathds{1}\otimes T_{R_2}^a\otimes  T_{ R_3a}\\&+\frac{\hbar^2}{4(z_2-z_3)^2}\big(\mathds{1}\otimes T^a_{R_2}T^b_{R_2}\otimes T_{R_3a}T_{R_3b} + \mathds{1}\otimes T^a_{R_2}T^b_{R_2}\otimes  T_{R_3b}T_{R_3a}\big)+\mathcal{O}(\hbar^3).
\end{aligned}
\end{equation}

We shall now make use of the identity
\begin{equation}\label{cyb}
\frac{[T^a_{R_1},T^b_{R_1}]\otimes T_{R_2a}\otimes T_{R_3b}}{(z_1-z_2)(z_1-z_3)} +\frac{T^a_{R_1}\otimes [T_{R_2a},T^b_{R_2}]\otimes T_{R_3b}}{(z_1-z_2)(z_2-z_3)} +\frac{T^a_{R_1}\otimes T^b_{R_2}\otimes [T_{R_3a},T_{R_3b}]}{(z_1-z_3)(z_2-z_3)} =0,
\end{equation}
which follows since the classical $r$-matrix, $r_{ij}=\frac{T^{a}_{R_i}\otimes T_{R_ja}}{z_i-z_j}$ ($i,j=1,2,3$, where $j>i$), obeys the classical Yang-Baxter equation.  Using \eqref{cyb}, %the second term in each parenthesis in 
the last three terms at order $\hbar^2$ of \eqref{threee} can be shown to be 
\begin{equation}
\begin{aligned}
&\frac{\hbar^2}{(z_1-z_2)(z_1-z_3)}T^a_{R_1}T^b_{R_1}\otimes T_{R_2a}\otimes T_{R_3b}+\frac{\hbar^2}{(z_1-z_2)(z_2-z_3)}T^a_{R_1}\otimes T_{R_2a}T_{R_2b}\otimes T^b_{R_3}  \\&+\frac{\hbar^2}{(z_1-z_3)(z_2-z_3)}T^a_{R_1}\otimes T^b_{R_2}\otimes T_{R_3a}T_{R_3b},
\end{aligned}
\end{equation}
whereby \eqref{threee} agrees with the LHS of \eqref{YB}. Alternatively, we can use \eqref{cyb} such that the last three terms at order $\hbar^2$ of \eqref{threee} are
\begin{equation}
\begin{aligned}
&\frac{\hbar^2}{(z_1-z_2)(z_1-z_3)}T^a_{R_1}T^b_{R_1}\otimes T_{R_2b}\otimes T_{R_3a}+\frac{\hbar^2}{(z_1-z_2)(z_2-z_3)}T^a_{R_1}\otimes T^b_{R_2}T_{R_2a}\otimes T_{R_3b}  \\&+\frac{\hbar^2}{(z_1-z_3)(z_2-z_3)}T^a_{R_1}\otimes T^b_{R_2}\otimes T_{R_3b}T_{R_3a},
\end{aligned}
\end{equation}
whereby \eqref{threee} agrees with the RHS of \eqref{YB}. Thus, the boundary six-point function \eqref{threee} is in agreement with the bulk correlation function of three Wilson lines up to order $\hbar^2$, modulo the framing anomaly. We expect that this  will hold at higher orders of $\hbar$ as well.
%\newline
% \newline
 %	\noindent{\textit{Acknowledgements}}
 \acknowledgments
 	
 	We would like to thank %Benjo Fraser, Thomas Mertens, 
 	Benjo Fraser, Lennart Schmidt, Meng-Chwan Tan, Junya Yagi, and Masahito Yamazaki for helpful discussions and suggestions, and the anonymous referee for helpful comments. %Content 
 	%Preliminary r
 	Results in 
 	%of 
 	this paper were presented at Kavli IPMU in May, 2019, and we thank the audience of this talk for feedback. We would also like to thank Kavli IPMU
 	%, where part of this work was performed, 
 	for hospitality. This work is supported 
 	%in part by the NUS FRC Tier 1 grant R-144-000-377-114 and 
 	by the MOE Tier 2 grant R-144-000-396-112.

\end{document}